\documentclass[sigconf, screen]{acmart}

\AtBeginDocument{%
  }

\setcopyright{none} 
\copyrightyear{}

\acmConference[arXiv]{arXiv Preprint}{2026}{}

\usepackage{tcolorbox}

\begin{document}
\begin{sloppypar}

\title{Loosely-Structured Software: Engineering Context, Structure, and Evolution Entropy in Runtime-Rewired Multi-Agent Systems}
\subtitle{}

\author{Weihao Zhang}
\affiliation{%
  \institution{The Hong Kong University of Science and Technology}
  \country{Hong Kong SAR, China} 
}

\author{Yitong Zhou}
\affiliation{%
  \institution{AI Chip Center for Emerging Smart Systems}
  \country{Hong Kong SAR, China}
}

\author{Huanyu Qu}
\affiliation{%
  \institution{University of Macau}
  \country{Macau SAR, China} 
}

\author{Hongyi Li}
\affiliation{%
  \institution{Tsinghua University}
  \country{China} 
}




\begin{abstract}

As LLM-based multi-agent systems (MAS) become more autonomous, their free-form interactions increasingly dominate system behavior. However, scaling the number of agents often amplifies context pressure, coordination errors, and system drift. It is well known that building robust MAS requires more than prompt tuning or increased model intelligence. It necessitates engineering discipline focused on architecture to manage complexity under uncertainty. We characterize agentic software by a core property: \emph{runtime generation and evolution under uncertainty}. Drawing upon and extending software engineering experience, especially object-oriented programming, this paper introduces \emph{Loosely-Structured Software (LSS)}, a new class of software systems that shifts the engineering focus from constructing deterministic logic to managing the runtime entropy generated by View-constructed programming, semantic-driven self-organization, and endogenous evolution.

To make this entropy governable, we introduce design principles under a three-layer engineering framework: \emph{View/Context Engineering} to manage the execution environment and maintain task-relevant Views, \emph{Structure Engineering} to organize dynamic binding over artifacts and agents, and \emph{Evolution Engineering} to govern the lifecycle of self-rewriting artifacts. Building on this framework, we develop LSS design patterns as semantic control blocks that stabilize fluid, inference-mediated interactions while preserving agent adaptability. Together, these abstractions improve the \emph{designability}, \emph{scalability}, and \emph{evolvability} of agentic infrastructure. We provide basic experimental validation of key mechanisms, demonstrating the effectiveness of LSS.

\end{abstract}



\maketitle

\section{Introduction}\label{sec:intro}

Large language model (LLM) multi-agent systems (MAS) are experiencing a ``Cambrian explosion''~\cite{wang2024survey,chase2022langchain,yang2023autogpt,wu2023autogen,hong2023metagpt,openclaw2026,nanoclaw2026}. Across academia and industry, teams are assembling systems that search, plan, code, and iterate, aiming to turn probabilistic models into reliable and productive infrastructure. Powered by semantic understanding and intelligent reasoning, LLM agents can synthesize information, propose actions, and coordinate across tools. Yet progress in MAS has also exposed a hard limit: capability does not scale linearly with the number of agents~\cite{kim2025towards}. In practice, larger teams often amplify errors and coordination overhead, and some claims argue that simpler agent architectures can better align with user intent~\cite{openai2025practicalagents}. A critical gap persists: while prototypes dazzle with emergent capabilities, production systems often hit a ceiling of complexity where additional agents add more failures than utility.

\begin{figure}[h]
  \centering
  \includegraphics[width=0.85\linewidth]{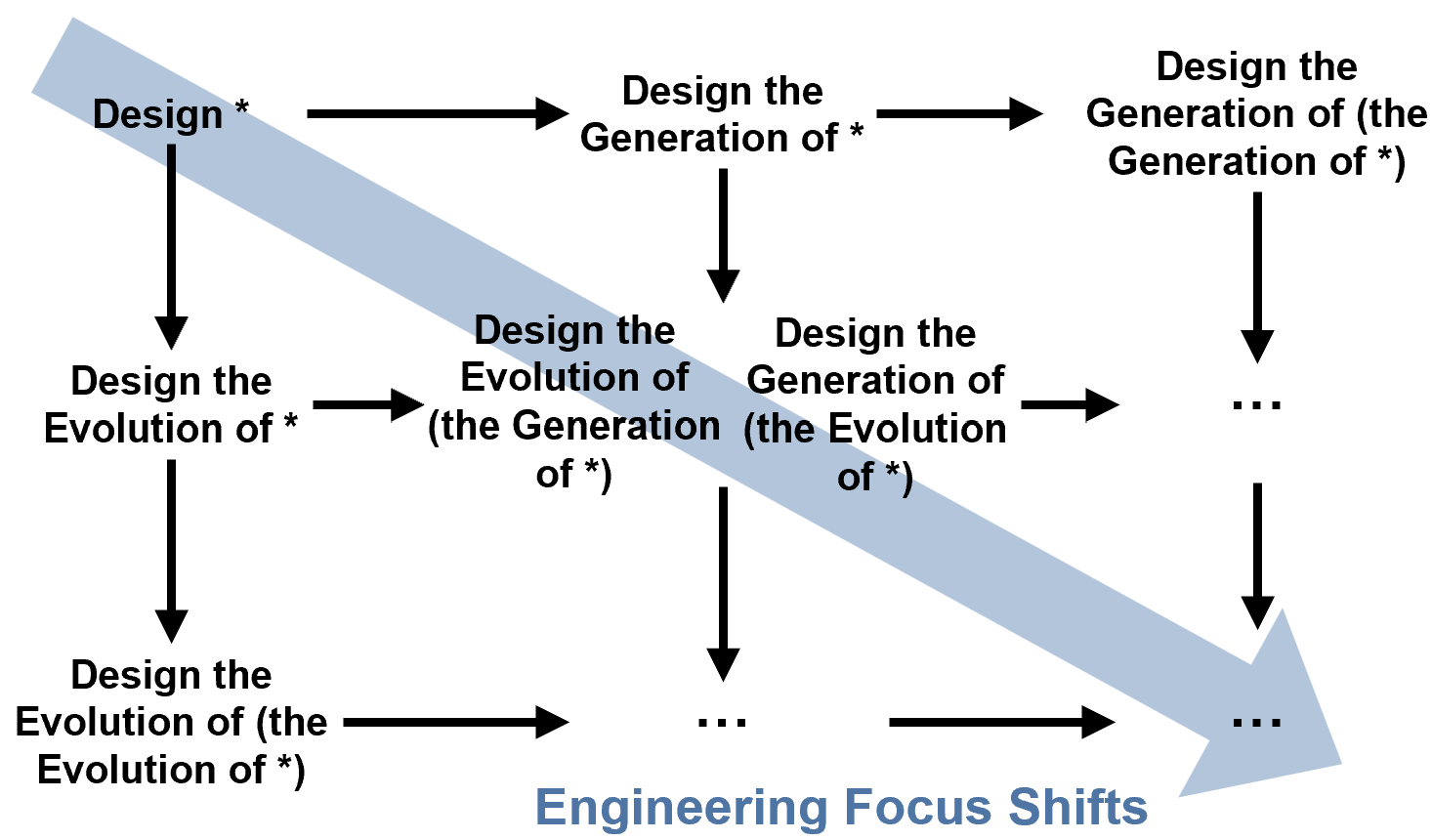}
  \caption{The engineering focus shifts from designing specific logical code to designing runtime generation and evolution.}
  \label{fig:lss-engineering-migration}
\end{figure}

Much of the current MAS literature focuses on building ``more intelligent'' orchestration frameworks~\cite{yao2023react,shinn2023reflexion,yao2023treeofthoughts,khattab2024dspy}. Increasingly, however, the dominant failures are not merely a matter of prompt tuning or base model intelligence, but an \emph{engineering problem} of architecture (recently emerging OpenClaw-/NanoClaw-like architectures~\cite{openclaw2026,nanoclaw2026} also reflect this trend): the discipline required to manage complexity, interfaces, and drift under uncertainty. Accordingly, this paper treats MAS as a new kind of software system and develops an engineering perspective that draws from, adapts, and extends conventional software engineering principles and patterns. MAS need not be treated as an opaque ``intelligence-on-intelligence'' black box; rather, they can be approached as software systems that can be designed, governed, tested, and evolved.

Engineering this new kind of software requires confronting its core property: \emph{runtime generation and evolution under uncertainty}. Classic software architecture assumes build-time modular decomposition and slow-changing boundaries. Agentic software violates these assumptions through three coupled ``physics'':

\begin{itemize}
    \item \textbf{View-Constructed Programming}: Unlike modular software where logic is encapsulated in code, an agent's effective program is determined by a step-specific \emph{View} constructed from global Artifact files (e.g., system prompts, skills, plans, tools, memories) and projected into its context window at runtime~\cite{mei2025contextengineering,packer2023memgpt,liu2023lostmiddlelanguagemodels}.
    \item \textbf{Runtime-Generated Structure, Abstraction, and Semantic Binding}: Unlike code-defined linking, component connections are formed dynamically via semantic understanding. The system's connectivity and abstraction can be generated on the fly via agent/tool selection, temporary team organization, and interfaces negotiated in natural language rather than fixed function signatures~\cite{wu2023autogen,hong2023metagpt,qian2024chatdev,karpas2022mrkl,schick2023toolformer}.
    \item \textbf{Endogenous Evolution}: The system's executable substrate (Artifact files that mediate behavior in-context) is itself rewritable by the system, and thus forms the foundation of adaptation and self-improvement~\cite{shinn2023reflexion,madaan2023selfrefine,wang2023voyager}.
\end{itemize}

This paper defines Loosely-Structured Software (LSS) as a software paradigm characterized by these properties. While many existing frameworks provide valuable structure through deterministic pipelines to ensure reliability (e.g., LangChain, AutoGPT)~\cite{chase2022langchain,yang2023autogpt}, LSS explores an alternative path by treating runtime variability as a first-class design dimension (Figure~\ref{fig:lss-engineering-migration}).

For engineers trained in traditional software development, this gap signals a deeper shift. Instead of writing deterministic logic behind stable interfaces, one must engineer the conditions under which agents construct Views, bind to capabilities, and revise their own Artifacts. A recurring challenge is deciding what to constrain and what to leave flexible. Traditional software design workflows favor stable abstractions (protocols, intermediate representations) and stable topologies (e.g., classic Object-Oriented Programming, OOP patterns). In LSS, however, abstractions and architecture are often unstable at design time and can be generated and revised at runtime. Moreover, because agents interpret the same Artifact through task-driven perspectives, the abstraction can be viewpoint-dependent rather than objective. Addressing these issues cannot rely solely on adding more higher-level wrappers.

\begin{figure*}[h]
  \centering
  \includegraphics[width=0.85\linewidth]{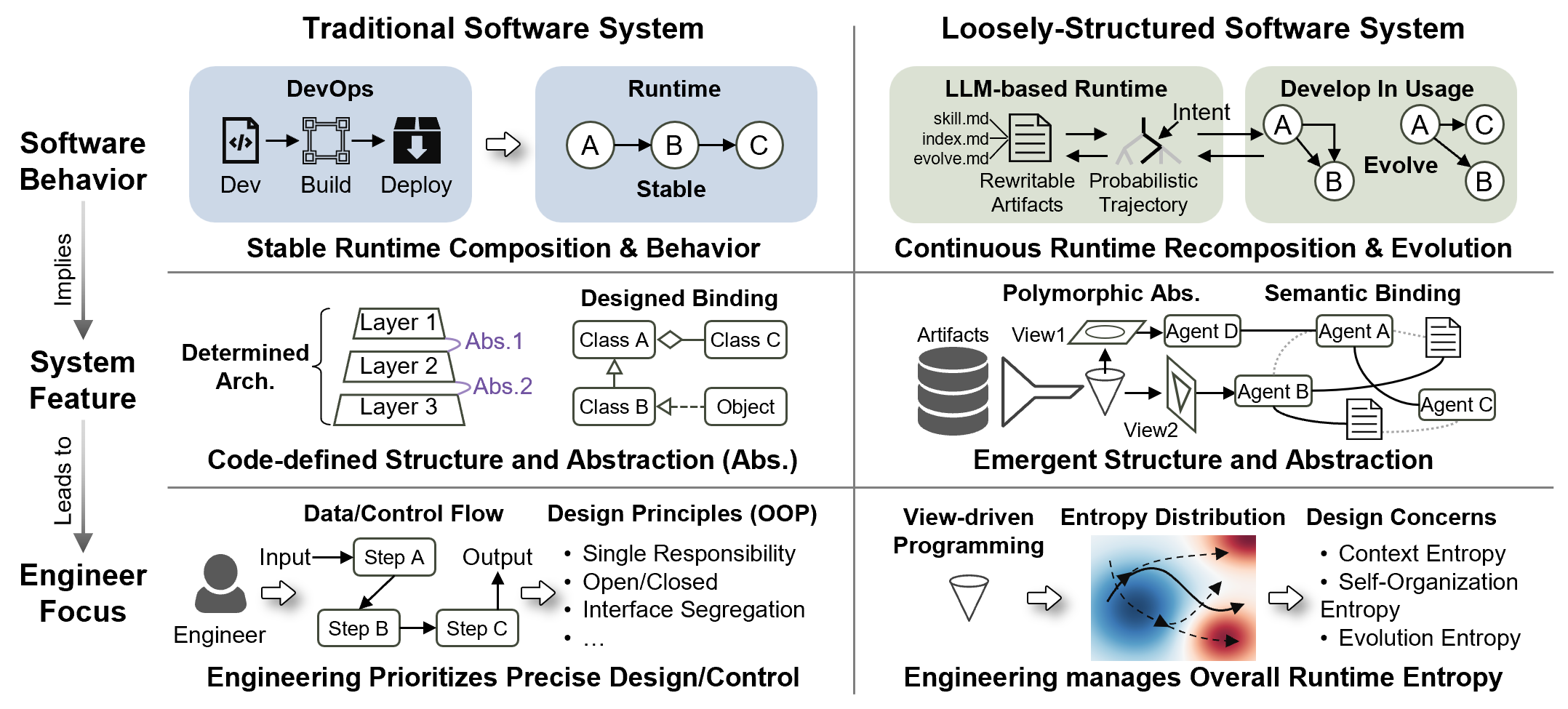}
  \caption{Loosely-Structured Software: a software system with unstable, runtime-generated structure \& abstraction and its dominant engineering focuses (View, Binding, Evolution).}
  \label{fig:lss-overview}
\end{figure*}

Beyond these design differences, practical MAS characteristics further complicate engineering: \textbf{(i) View-boundedness} (context budgets make context pollution a main failure source)~\cite{liu2023lostmiddlelanguagemodels,mei2025contextengineering,greshake2023indirectprompt}, \textbf{(ii) collaboration cognitive gaps} (agents often lack explicit supply--demand contracts, so coordination devolves into semantic probing)~\cite{wu2023autogen,qian2024chatdev,a2a2025spec}, and \textbf{(iii) persistent-memory and Artifact maintenance} (large, evolving stores of skills, traces, and long-term memories accumulate operational debt and require coherent management)~\cite{packer2023memgpt,zhong2023memorybank,xu2025amem}. Together, these characteristics drive the poor scalability observed in practice: as more agents are added, the system bears greater ``context pressure'' and a growing communication tax without proportional gains in task success.

Figure~\ref{fig:lss-overview} summarizes the key differences between LSS and traditional software in both behavior and system features. These differences shift the engineering objective from ``constructing deterministic logic'' to ``managing runtime entropy distribution'': guiding the growth of the system itself. We introduce a three-layer framework that systematizes this entropy management: \textbf{Layer 1 (View/Context Engineering)} manages the execution environment and attention span (taming Context Entropy); \textbf{Layer 2 (Structure Engineering)} manages the organization of artifacts and agents to enable dynamic capability discovery and binding (taming Self-Organization Entropy); and \textbf{Layer 3 (Evolution Engineering)} manages the lifecycle of self-rewriting artifacts (taming Evolutionary Entropy). 

Building on this framework, we also adapt reusable software design wisdom---especially from OOP~\cite{gamma1994designpatterns}---and intellectualized it as dynamic engineering mechanisms for regulating information flow and distribution. Unlike OOP design patterns that typically map to static class hierarchies and compile-time relationships, \emph{LSS Design Patterns} (e.g., \emph{Semantic Router}, \emph{Mediator}) operate as semantic control blocks that stabilize fluid, inference-mediated interactions. These patterns transform structural uncertainty into a feature that increases \textbf{designability}---ensuring that an intuitively designed agent organization yields expected behaviors; \textbf{scalability}---where ``more agents'' translates to higher task completion rather than increased noise; and \textbf{evolvability}---where the system can adapt to task environments and grow its own structure and capabilities through managed self-rewriting. 

This paper makes the following contributions:
\begin{itemize}
    \item We define \textbf{Loosely-Structured Software (LSS)} as a distinct paradigm, contrasting its dynamic ``physics'' with traditional static software engineering.
    \item We propose a \textbf{Three-Layer Framework} (View, Structure, Evolution) that systematizes the management of agent context, self-organization, and evolution entropy respectively.
    \item We formalize some reusable \textbf{Design Principles} and \textbf{Design Patterns} that provide concrete mechanisms for stabilizing fluid interactions, and we discuss how these logical patterns can be mapped into physical agent realizations through practical design strategies.
    \item We demonstrate through \textbf{Workflow Examples} how LSS principles solve common failure modes like context overflow and binding hallucination with multi-agent evaluation.
\end{itemize}

This paper is intended to be complementary to existing MAS research rather than a replacement. Some of the design principles and patterns discussed in this paper build on established ideas in the MAS and agentic-systems literature and practice. We reframe and extend them through a software engineering lens to recapture design intuitions. This also implies the limitations of this work. The goal of this paper is to provide higher-level design language, not an exhaustive catalog of implementation techniques or a detailed treatment of the technical pitfalls of each pattern across frameworks and deployments. We present experimental demonstrations in the Workflow Examples to illustrate how the proposed principles and patterns address common architectural concerns, but a comprehensive pattern-by-pattern implementation is beyond the scope of this paper.

\section{Background and Related Work}\label{sec:background}

\subsection{Basic Concepts of Multi-Agent Systems}
Multi-agent systems (MAS) are composed of autonomous entities that interact to solve problems beyond the capabilities of individual agents. In contemporary LLM-based systems, an agent is commonly modeled as an \emph{LLM-centered control loop} equipped with four key modules: \textbf{Profile}, \textbf{Memory}, \textbf{Planning}, and \textbf{Action}~\cite{wang2024survey}. This abstraction is useful because it separates (i) \emph{what the agent is supposed to be} (role and constraints), (ii) \emph{what it knows} (task-relevant context), (iii) \emph{how it decides} (goal decomposition and selection), and (iv) \emph{what it can do} (tool- and environment-facing operations).
\begin{itemize}
    \item \textbf{Profile}: Defines the role, persona, and constraints of the agent, often encoded in system prompts.
    \item \textbf{Memory}: Stores interaction history and knowledge, ranging from short-term working context to long-term vector stores or knowledge graphs.
    \item \textbf{Planning}: Decomposes complex user goals into executable subtasks and selects next steps under uncertainty (e.g., ReAct~\cite{yao2023react}).
    \item \textbf{Action}: Enables the agent to perceive and affect its environment via external tools and APIs, where tool schemas, permissions, and side effects become part of the execution boundary.
\end{itemize}

In production stacks, these conceptual modules map to concrete architectural components: \textbf{agents} (execution units), \textbf{tools} (external services), and \textbf{skills} (reusable procedures or playbooks encoded as prompts and structured steps).

Recent interoperability protocols further clarify these surfaces. The Model Context Protocol (MCP) standardizes how an agent host connects to external tool and data servers through a discoverable, typed interface, making ``tool access'' a portable capability across applications~\cite{hou2025model,mcp2025spec}. Complementarily, Agent2Agent (A2A) standardizes how agents advertise capabilities and coordinate work across boundaries (e.g., long-running tasks and modality negotiation), making ``agent access'' portable across heterogeneous agentic applications~\cite{a2a2025spec}. Together, MCP and A2A expose a layered interoperability stack, while skills remain lightweight, in-context procedures that guide how tools are invoked and how agents collaborate.

A practical distinction in this architecture is the trade-off between stability and agility, particularly between \emph{tool interfaces} and \emph{skill procedures}. Tools stabilize delivery via explicit schemas and implementations, while skills are agile, natural-language procedures executed in-context and frequently regenerated.

\subsection{Related Work}
LSS differentiates itself by addressing the \emph{architectural mismatch} in current agentic engineering: the attempt to constrain probabilistic, fluid intelligence within static, deterministic software patterns. We categorize existing work through this lens.

\textbf{Loose Coupling Software.}
Loose coupling is an established concept in software engineering, spanning modular decomposition and information hiding~\cite{parnas1972criteria} through Service-Oriented Architecture (SOA) and microservices~\cite{dragoni2017microservices}. It aims to reduce change propagation by minimizing shared assumptions, narrowing dependency surfaces, and mediating interactions through stable abstractions; classic guidance frames this in terms of cohesion--coupling trade-offs~\cite{stevens1974structureddesign}. In distributed systems, loose coupling is also discussed via interaction styles that decouple components in time and space (e.g., publish/subscribe), reducing direct knowledge and synchronization requirements between producers and consumers.

In many practical stacks, loose coupling is operationalized through explicit, versioned interface contracts (e.g., IDLs, schemas, REST-style APIs). Such contracts stabilize syntactic interoperability, yet they also introduce coupling through shared definitions and compatibility management, especially under ongoing API evolution~\cite{zhong2023microservicecoupling,lercher2024microserviceapievolution}. LSS emphasizes decoupling at the semantic layer: components coordinate through natural-language intents and runtime interpretation, enabling probabilistic \emph{Semantic Binding} beyond fixed contracts.

\textbf{Context Engineering.}
Research in retrieval-augmented generation (RAG)~\cite{lewis2020rag} and long-context management (e.g., MemGPT~\cite{packer2023memgpt}) studies how to assemble task-relevant information for LLMs under context-window constraints, often discussed as \emph{context engineering} in recent surveys~\cite{mei2025contextengineering}. This includes retrieval and reranking, context compaction (summarization/compression), and structured memory mechanisms, as well as variants that interleave retrieval and generation with self-critique~\cite{asai2023self}. It is also motivated by empirical findings on long-context utilization (e.g., attention degradation and ``lost in the middle'' effects), which influence how context should be ordered, scoped, and refreshed during multi-step execution~\cite{liu2023lostmiddlelanguagemodels}.

\textbf{Orchestration Frameworks.}
Frameworks such as LangChain~\cite{chase2022langchain}, AutoGen~\cite{wu2023autogen}, and MetaGPT~\cite{hong2023metagpt} provide reusable primitives for defining agents, tools, and message passing, while DSPy~\cite{khattab2024dspy} explores declarative compilation and optimization of prompt pipelines. This line of work motivates practical design choices around interaction topology (e.g., graphs, pipelines, routers) and the mechanics of routing, delegation, and tool invocation.

\textbf{Autonomous Agents.}
Autonomous agent systems such as AutoGPT~\cite{yang2023autogpt} and BabyAGI explore recursive self-prompting and tool-augmented task execution, while later work such as Voyager~\cite{wang2023voyager} and Generative Agents~\cite{park2023generativeagents} studies long-horizon learning and open-ended social simulation. Across these systems, a recurring engineering concern is how to represent and update agent state (e.g., memories and skill libraries) and how to incorporate verification and rollback mechanisms into long-running loops.

\textbf{Harness Engineering.}
The recent notion of \emph{harness engineering} frames agentic software development as the design of the surrounding environment that makes agents reliable: repository-local documentation, deterministic guardrails (linters, structural tests), automated feedback loops, and runtime telemetry that agents can inspect while iterating. OpenAI reports using such a harness to build and maintain a large codebase with Codex agents, with the harness spanning tests, integration, documentation, and observability~\cite{openai2026harnessengineering}. Fowler further discusses harnesses as a practical vocabulary for the tooling and practices that keep agents ``in check'' while enabling sustained iteration at scale~\cite{fowler2026harnessengineering}.

\textbf{Summary.}
While existing work excels at providing the \emph{mechanisms} (how to call an LLM, how to store vectors), LSS seeks to synthesize these diverse engineering practices into a unified architectural vocabulary. It aims to provide a complementary layer of formalization that helps developers reason about system-level entropy.

\section{Definition of Loosely-Structured Software}\label{sec:definition}

In this section, we formalize the concept of \textit{Loosely-Structured Software (LSS)} to distinguish it from traditional software. We first identify four conceptual features that characterize LSS systems. Based on these features, we then introduce a system model that captures the essential runtime dynamics of such systems. Finally, we outline a three-layer conceptual framework that organizes the core engineering problems of LSS.

\subsection{Conceptual Features}
We describe a system as \textit{Loosely-Structured Software (LSS)} when it exhibits the following four conceptual features:

\textbf{View-Constructed Programming.} 
The system controls execution by repeatedly constructing a step-specific \textit{View} from global information. Under context window constraints, it selects, filters, and projects the information scope for the current step, rather than operating directly on the full dataset or statically encoding a fixed information scope per step (such as explicitly restricting the scope of a variable in the code in traditional programming).

\textbf{Runtime Semantic Binding.} 
The invocation relationships between artifacts are not determined solely by hardcoded linking (such as functional call, inheritance, composition, or message passing). Instead, the system employs \textit{Semantic Routing} to dynamically bind agents and intents to artifacts and construct execution and information flow paths at runtime, achieving pure runtime-defined binding of targets.

\textbf{Endogenous Evolution.} 
Core logic is encoded in readable and rewritable \textit{Artifacts} (e.g., skill files) rather than immutable code. The system can programmatically read, analyze, and \textit{Rewrite} these Artifacts at runtime, enabling behavior change by modifying the Artifacts themselves.

\textbf{Dynamic Abstraction Conversion.} 
Abstraction boundaries are not statically fixed. The same artifact can be used in different roles across Views---serving as an interface definition, an intermediate contract, or concrete implementation for analysis—with abstractions generated and reinterpreted on demand as the execution context shifts.

\subsection{System Model: Runtime Elements and Primitives}

Agent frameworks often model an agent as a bundle of files, with the LLM abstracted as a background reasoning engine. In practical agentic execution, behavior is trajectory-dependent: what the agent can do at step $t$ is shaped by the accumulated interaction history and, critically, by the step-specific information that the system constructs from files and injects into the model. In LSS, an agent's capability, stability, and identity are therefore realized via step-by-step construction rather than being fixed as a single predeclared program. To formalize this runtime, we describe LSS using four fundamental elements: \textit{Intent}, \textit{Global Artifacts}, \textit{View}, and \textit{Output}.

\begin{figure}[t]
  \centering
  \includegraphics[width=1\linewidth]{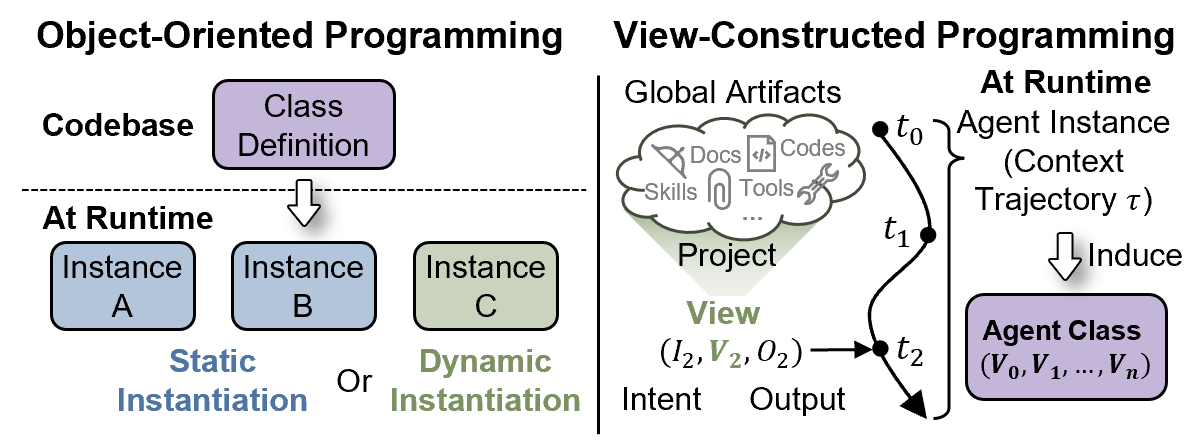}
  \caption{Comparison of Architectural Paradigms: Traditional Software vs. LSS. Left: Traditional software relies on compile-time program structure to constrain runtime behavior. Right: LSS makes behavior depend on the runtime trajectory of Views and Artifacts, which induces the effective agent configuration over time.}
  \label{fig:lss-vs-trad}
\end{figure}

\textbf{1. Four Core Runtime Elements.} 
For formalization, we model the LSS from the perspective of ongoing interaction trajectories, where $t$ denotes the discrete interaction step. At any step $t$, the system dynamics are governed by:
\begin{itemize}
  \item \textbf{Artifacts ($A_t$):} The global collection of all persistent artifacts at step $t$. The LLM is abstracted away as part of the default infrastructure. We view LSS as a collection of files; we call files that shape the system's capabilities and behavior \textit{artifacts}. Artifacts can include reusable prompts (e.g., system prompts, \texttt{agent.md}), skills (e.g., \texttt{skill.md}), plans, code, tool registries (e.g., MCP), contracts (e.g., A2A), traces, documents, databases, and memories used by some MAS frameworks~\cite{openclaw2026}. They also include artifacts introduced in this paper (e.g., \texttt{index.md} for defining the file system, \texttt{contract.md} and \texttt{team.md} for agent collaboration, \texttt{fork.md} for agentic inheritance, \texttt{evolve.md} for system evolution, \texttt{lens.md} and \texttt{route.md} for semantic binding, \texttt{task.md} in the automatic task CI/CD). Artifacts can be viewed as the system's ``hard drive'' at step $t$. They are indexed by $t$ because the system can rewrite artifacts, causing the global capability space to evolve over time.
  \item \textbf{Intent ($I_t$):} The explicit query or driving force of the current computation step (e.g., an external user instruction, a subtask invoked from another agent, or a self-driven intent). It acts as the semantic input of the "Agent program".
  \item \textbf{View ($V_t$):} The specific, transient projection dynamically assembled from $A_t$ and injected into the LLM's context window. It acts as the system's ``active RAM.'' Logically, $V_t$ is distinct from $I_t$: while $I_t$ is the objective to be solved, $V_t$ is the step-specific execution scope required to solve it (a step-level executable program). However, the boundary between them can be subjective, depending on the agent's perspective when executing the task.
  \item \textbf{Output ($O_t$):} The contents or actions generated by the LLM and the automatic environmental feedback (e.g., tool execution results). It closes one interaction step.
\end{itemize}

\textbf{2. Architectural Inversion: Instances Inducing Classes.} 
By tracking these four elements over time (Figure \ref{fig:lss-vs-trad}), we reveal a fundamental architectural inversion between traditional object-oriented software and LSS. In traditional OOP, a developer usually statically defines a \textit{Class} in the codebase, which then explicitly spawns execution \textit{Instances} at runtime. In LSS, the relationship can be inverted. First, multiple interaction steps define an \textbf{Agent Instance} as the complete context trajectory ($\tau$) of its execution up to step $n$:
\begin{equation}
\text{Agent Instance} (\tau_n) \equiv \{(V_0, I_0, O_0), (V_1, I_1, O_1), \dots, (V_n, I_n, O_n)\}
\end{equation}
This trajectory captures the specific, ongoing physical execution of an agent solving a problem.

We ignore cases where an agent framework fine-tunes LLM weights. LSS conceptually abstracts the LLM as a general-purpose, uniform reasoning engine whose behavior is conditioned by the runtime View sequence. Consequently, the structural identity, capabilities, and cognitive boundaries of the agent---its \textbf{Agent Class}---are not fixed upfront. Instead, the Agent Class is defined by the View sequence it is exposed to, which is dynamically \textit{induced} by the Agent Instance:
\begin{equation}
\text{Agent Class} \equiv \{V_0, V_1, \dots, V_n\}
\end{equation}
This suggests a shift in perspective within the LSS paradigm: where the \textbf{'Agent Class' can be viewed as an emergent property induced by its runtime Context trajectory.} Even if two agent instances start with the identical initial profile in $A_t$, a divergence in their runtime View sequences (e.g., one retrieving clean API schemas and the other retrieving polluted conversational noise) means they may mutate into entirely different Agent Classes mid-execution. Some rule-based software can also generate types or behaviors at runtime via mechanisms such as reflection and metaprogramming~\cite{kiczales1991amop,marr2015zerooverhead}, but most mainstream OOP design patterns are not built on the assumption that class structure is routinely generated at runtime. In LSS, by contrast, this kind of runtime-induced structure is a core characteristic.\footnote{To clarify the conceptual shift, we contrast LSS with a stylized model of traditional software—acknowledging that real-world systems often exhibit hybrid behaviors.}

\begin{center}
\begin{tcolorbox}[colback=gray!10, colframe=black!50, arc=2mm, boxrule=1pt, width=0.95\linewidth]
\textbf{\textit{Comparison:}}
\begin{itemize}
    \item \emph{\textbf{Traditional software:} Class defines instances.} 
    \item \emph{\textbf{LSS:} Agent instances define class.}
\end{itemize}
\end{tcolorbox}
\end{center}

\textbf{3. The LSS Execution Cycle.} 
Based on this formalization, the continuous execution cycle of an LSS system transitions the state from step $t$ to $t+1$ and can be modeled via a set of primitives, which dynamically operate over the global Artifacts, the current Intent, and the historical context (let $\tau_{<t} = \{(V_0, I_0, O_0), \dots, (V_{t-1}, I_{t-1}, O_{t-1})\}$ denote the trajectory up to step $t{-}1$):

\begin{enumerate}
    \item \textbf{Perception (View Construction):} $V_t = \text{Project}(A_t, I_t, \tau_{<t})$. The $\text{Project}$ primitive assembles a step-specific View through a coupled selection process: $I_t$ and $\tau_{<t}$ jointly shape what gets retrieved and prioritized from $A_t$, and what ultimately gets included in the resulting $V_t$. Note that every procedure in LSS can be semantic: $\text{Project}$ not only selects relevant Artifacts, but can also transform them (e.g., compressing, filtering, or merging information) based on the Intent. In practice, $\text{Project}$ typically requires only a subset of, or a compressed representation of, $\tau_{<t}$.
    
    \item \textbf{Decision \& Action (Execution):} $O_t = \text{Execute}(\text{LLM}(V_t, I_t, \tau_{<t}))$. The underlying LLM takes the constructed View $V_t$, the explicit objective $I_t$, and the historical context $\tau_{<t}$ as input. The $\text{Execute}$ primitive resolves and performs the agent's inference and carries out any accompanying operations (e.g., executing a Python script, calling an API) to produce the output result $O_t$.
    
    \item \textbf{Evolution (Artifact Update):} $A_{t+1} = \text{Update}(A_t, \tau_{t})$. The $\text{Update}$ primitive uses the current trajectory $\tau_t$ (including the latest $(V_t, I_t, O_t)$) to modify the system's Artifacts (e.g., rewriting a skill file or recording a distilled trace), evolving the global artifact space $A$ for the future.
    
    \item \textbf{Propagation (Intent Formulation):} $I_{t+1} = \text{Formulate}(\tau_{t})$. The $\text{Formulate}$ primitive analyzes the current context to derive or generate explicit Intents for either itself or other agents. This entails breaking down a complex result into a new subtask, issuing a self-correction incentive, or formulating a query for another agent.
\end{enumerate}

\begin{figure}[h]
  \centering
  \includegraphics[width=1\linewidth]{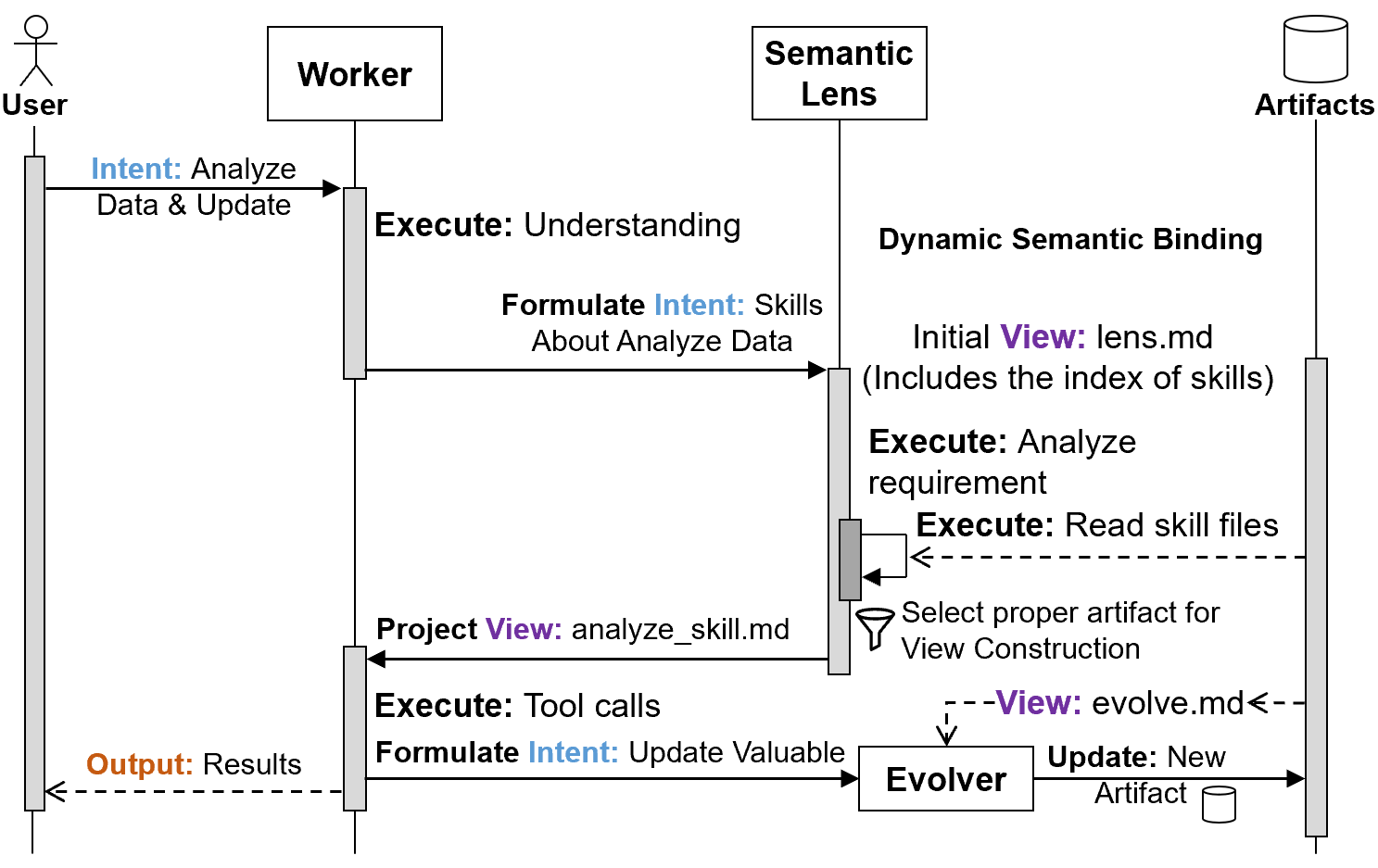}
  \caption{Multi-agent interaction in LSS, represented using UML-style sequence diagram notation. The diagram illustrates how the primitives decompose the loosely-structured runtime.}
  \label{fig:lss-sequence}
\end{figure}

Crucially, in an LSS architecture, these primitives do not have to be executed sequentially by a single monolithic entity. For instance, in Figure \ref{fig:lss-sequence}, a ``Lens'' might execute the $\text{Project}$ primitive to construct a $V_t$, which is then handed over to a ``Worker'' to $\text{Execute}$. Similarly, the $I_{t+1}$ generated by the $\text{Formulate}$ primitive can be targeted at an entirely different agent. This also implies that the same information can play different roles for different agents: what is an Output for one agent can become part of the View for another agent (e.g., the Router's $O_t$ becomes the Worker's $V_t$). In Section~\ref{sec:cohesion}, we detail the mapping between these roles and the concrete agents in our implementation. Multiple roles can be embedded into one agent based on the Semantic Cohesion Principle. Before introducing that, we can view each role as a logically isolated agent for simplicity.

\begin{figure}[h]
  \centering
  \includegraphics[width=0.95\linewidth]{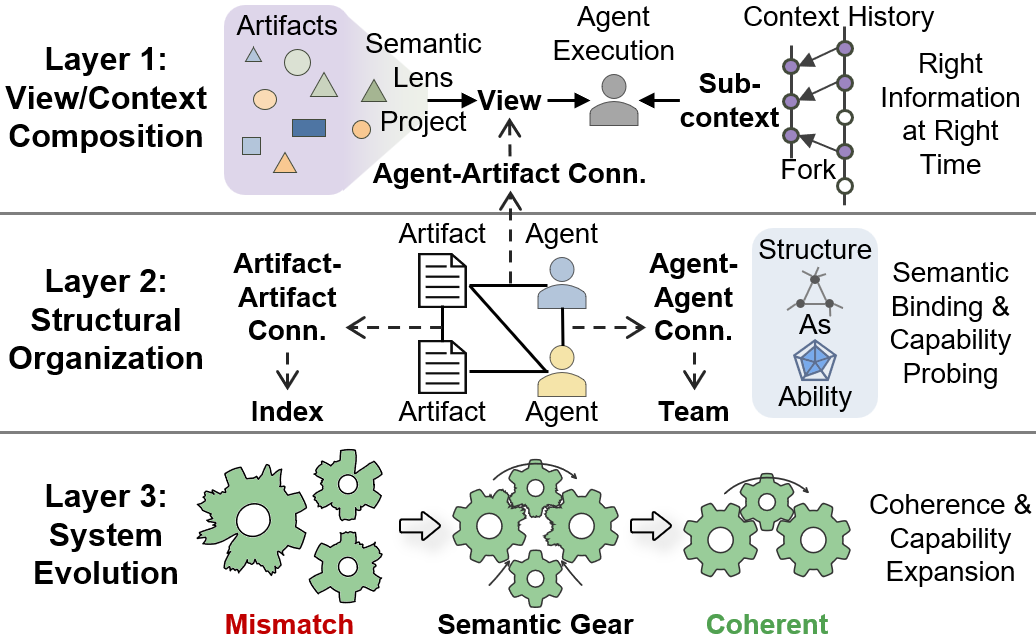}
  \caption{Three-layer engineering framework for LSS: View--Context, Structure--Capability, and Evolution--Adaptation.}
  \label{fig:lss-three-layers}
\end{figure}

Together, these runtime elements and primitives make the structure and abstraction dynamics of LSS universal. Unlike traditional software, where developers mostly precisely predefine control logic and information flow, LSS accepts runtime uncertainty induced by View construction, semantic binding, and Artifact evolution. Engineering LSS therefore shifts from exact control to semantic entropy management. Concretely, we identify three forms of runtime entropy: \textbf{Context Entropy, Section~\ref{sec:layer1}}, \textbf{Self-Organization Entropy, Section~\ref{sec:layer2}}, and \textbf{Evolutionary Entropy, Section~\ref{sec:layer3}}. It can be imagined that if LSS owns thousands of artifacts or hundreds of agents, the runtime uncertainty will be extremely large and lead to system failure~\cite{kim2025towards}. Figure \ref{fig:lss-three-layers} organizes these challenges into three layers (View--Context, Structure--Capability, and Evolution--Adaptation). The following sections discuss possible engineering principles and patterns for taming each entropy respectively.

\section{View/Context Engineering: Taming Context Entropy}\label{sec:layer1}

The first layer concerns the immediate execution environment, shifting the engineering focus from designing \emph{Control Flow} to designing \emph{Context Flow}. In traditional software, the runtime environment (stack, heap, variable scope) is managed by the OS and compiler. In LSS, the runtime environment for each reasoning step is the information projected into the model's View: whatever enters the View participates in execution by activating capabilities (tools and skills), constraining behavior (policies), and biasing decisions (examples and retrieved memories). This is \emph{View-Constructed Programming}: rather than executing a fixed program against different inputs, the system repeatedly assembles a step-specific ``executable slice'' from a large Artifact pool (prompts, skills, documents, traces) and then runs the model within that slice~\cite{brown2020fewshot,reynolds2021promptprogramming,lewis2020rag,packer2023memgpt}. The engineering question becomes how to govern View construction so produced Views are stable, economical, and aligned with the step Intent.

\textbf{Challenge: Context Entropy \& Two-Sided Failure.}
We define \textbf{Context Entropy} as the instability introduced by the gap between the \emph{actual} View projected to the agent and the \emph{ideally most helpful} View for the current step. View construction makes information boundaries highly dynamic (the textual execution scope can be assembled and recombined arbitrarily), and long contexts further degrade the model's ability to reliably use evidence. This entropy manifests as a two-sided failure mode:
\begin{itemize}
    \item \textbf{Excess (Context Pollution):} Providing too much information introduces irrelevant noise and conflicting constraints, leading to attention dilution~\cite{greshake2023indirectprompt,wen2025instructdetector}.
    \item \textbf{Deficiency (Context Starvation):} Providing too little information causes the absence of critical constraints or local state, leading to reasoning breaks and execution errors~\cite{lewis2020rag,asai2023self}.
\end{itemize}
Both failure modes are systematic rather than incidental. Retrieval may surface stale or weakly relevant memories; an upstream agent may over-serialize its chain-of-thought into downstream context~\cite{wei2022chainofthought,wang2022selfconsistency}; and tool outputs may encode assumptions that are each locally reasonable but jointly inconsistent. Because the View is assembled at runtime, small inclusion/omission changes can flip behavior, causing capabilities to appear ``non-deterministic'' even when the model is unchanged. The core difficulty is thus not merely context window size, but enforcing step-level View governance---what is revealed, when, and under which contracts---so the executed View stays close to the ideal View under a bounded cognitive budget.

\subsection{Design Principles}
\textbf{Progressive Disclosure.}
To reduce Context Entropy, View exposure should be staged based on the agent's confidence and the step Intent:
\begin{itemize}
    \item \textbf{Minimal Sufficient:} Default to projecting only what is necessary for the current step; keep everything else out to reduce accidental coupling and instruction interference.
    \item \textbf{Adaptive Context Expansion:} When unconfidence rises, expand the View with additional semantically relevant evidence to increase task success probability~\cite{asai2023self}.
    \item \textbf{Context Backpressure:} Use observable context pressure (e.g., token budget, ambiguity signals) to throttle exposure or compress and summarize under high pressure. Allow more evidence under low pressure~\cite{packer2023memgpt}.
\end{itemize}
This is the most widely adopted context-engineering principle in current MAS frameworks~\cite{mei2025contextengineering,openai2025practicalagents}. In practice, an \emph{agent skill} is already a concrete embodiment of Progressive Disclosure: a skill packages a task into step-wise, View-sized instructions and constraints that can be selectively loaded as execution progresses, rather than dumping the entire Artifact pool into the View at once.

\textbf{Step-level Customization.}
To reduce long-run cross-contamination, context engineering can also be customized at the granularity of each step as an optional governance tactic:
\begin{itemize}
    \item \textbf{Context Branching \& Stitching:} For complex sub-tasks, \texttt{fork} multiple clean sub-contexts (i.e., multiple reasoning branches) to isolate local reasoning traces and avoid contaminating the main thread, and only \texttt{stitch} a distilled outcome back~\cite{yao2023treeofthoughts,wang2022selfconsistency}.
    \item \textbf{Context Isolation:} For each step, construct a temporary sub-context by selecting a subset of the trajectory $\tau_{<t}$ that best matches the current Intent, then merge the step outcome into the main context.
\end{itemize}
Note that many agent frameworks already provide context compression and cleanup mechanisms that can serve as building blocks for these operations.

\begin{center}
\begin{tcolorbox}[colback=gray!10, colframe=black!50, arc=2mm, boxrule=1pt, width=0.95\linewidth]
\textbf{\textit{Comparison:}}
\begin{itemize}
    \item \emph{\textbf{Traditional software:} Variable scope is largely fixed at runtime.} 
    \item \emph{\textbf{LSS:} Effective scope (the step View and its carried history) is dynamically constructed at runtime.}
\end{itemize}
\end{tcolorbox}
\end{center}

\begin{figure}[h]
  \centering
  \IfFileExists{Figure/Figure_mediator_a2a_sequence.png}{\includegraphics[width=1\linewidth]{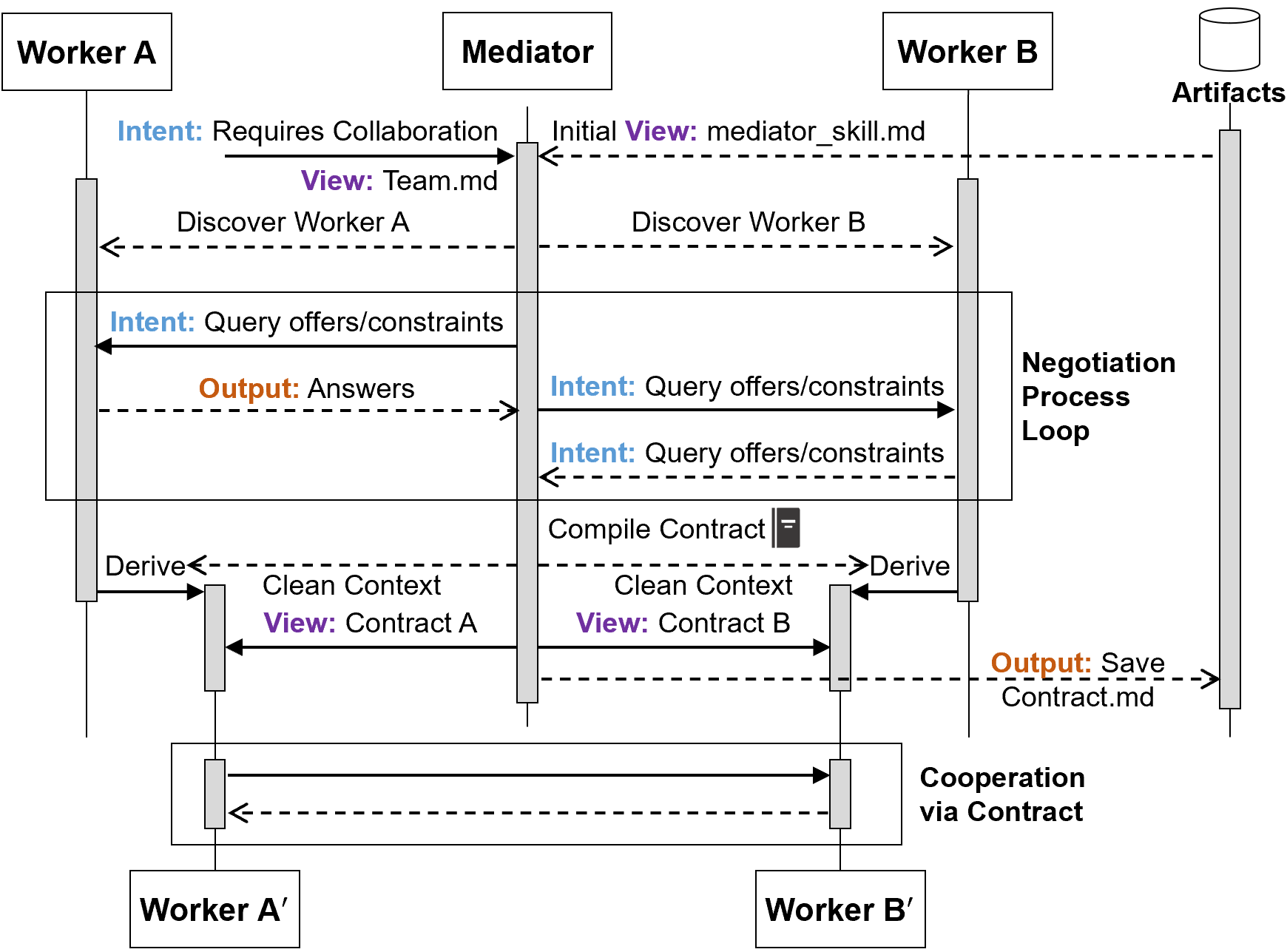}}{\fbox{\parbox{0.95\linewidth}{\centering Placeholder: \texttt{Figure/Figure\_mediator\_a2a\_sequence.png}}}}
  \caption{Sequence diagram of a Mediator constructing a task-specific A2A protocol between two agents.}
  \label{fig:mediator-a2a-seq}
\end{figure}

\subsection{Design Patterns}

\textbf{Semantic Lens.}
\emph{Intent.} Retrieve and compose the right artifacts/information on demand.
\emph{Mechanism.} When the global Artifact pool becomes too large for a worker agent to reliably self-select, even exposing only artifact metadata (e.g., thousands of skills) can impose significant context pressure, and metadata alone may be insufficient to choose reliably among many near-duplicate skills without reading their bodies. In this case, a dedicated agent (or module) can implement the \textbf{Project} runtime primitive: given the step Intent and trajectory, it retrieves and assembles a compact View from the relevant artifacts (often starting minimal and expanding when needed), thereby operationalizing Progressive Disclosure in View construction. Because a Lens can repeatedly generate an agent's step Views along a trajectory, it can also be viewed as an \emph{Agent Class Generator} that induces different worker classes via different View policies.

\textbf{Context Curator.}
\emph{Intent.} Keep agents' execution history more usable by distillation and compression.
\emph{Mechanism.} The Curator distills and compresses an agent's accumulated context, then produces a step-level temporary sub-context for the next operation. Since an Agent Class can be characterized by its View trajectory, the sub-context effectively induces a temporary ``narrower subclass'' for that step. We revisit and generalize this mechanism as \emph{agentic inheritance} in the next section, and the Context Curator can be viewed as an \emph{Inheritance Generator}.

\textbf{Mediator.}
\emph{Intent.} Enable collaboration without mutual View pollution.
\emph{Mechanism.} In LSS, a Mediator can be seen as dynamically constructing a \emph{task-specific} A2A-style collaboration contract: it concentrates the negotiation in its own View, distills the result into a clean contract, and then delivers only the protocolized outcome to the Workers. Concretely, the coordination unfolds as: (i) the Mediator receives an Intent that requires collaboration among agents; (ii) the Mediator iteratively negotiates and compiles an explicit contract (e.g., roles, I/O schema, state commitments, and allowed side effects); (iii) instead of continuing with the original agents (whose Views now contain bargaining traces), the Mediator forks two derived, clean-scoped child agents that inherit only the minimal necessary trajectory and the finalized contract in their context trajectories; (iv) the child agents execute tasks under contract-aligned Views. Figure~\ref{fig:mediator-a2a-seq} illustrates this flow, and reusable protocols can be persisted as \texttt{contract.md} artifacts for future reuse.

\textbf{End Criteria.}
\emph{Intent.} Decide when an agent instance has finished its responsibility and can be safely retired.
\emph{Mechanism.} In LSS, many agent instances are intentionally ephemeral: they are terminated once a transient task is completed. End criteria can attach a set of completion predicates to an agent (e.g., required outputs produced, invariants satisfied, budget exhausted, receipt of a specific signal from the user or another agent, etc.). Once the criteria are met, the agent can be terminated, and optional termination hooks can be triggered to finalize side effects (e.g., Claude Code hooks): for example, distilling and returning a summary to other agents, or archiving reusable artifacts.

\section{Structure Engineering: Taming Self-Organization Entropy}\label{sec:layer2}

This layer focuses on \emph{Structure Engineering}: governing runtime bindings so that artifact and agent topologies form in a task-appropriate way. Unlike traditional software where most dependency edges are fixed at compile time, ``structure'' in LSS is not a static module graph; it is the \textbf{binding topology} that emerges as the system repeatedly performs runtime semantic reasoning. Each binding decision---routing a message to an agent~\cite{wu2023autogen,hong2023metagpt,qian2024chatdev}, selecting an artifact, invoking a skill, or using a tool~\cite{yao2023react,schick2023toolformer}---implicitly creates a dependency edge for that step. These edges accumulate into a topology: which capabilities are frequently composed, which agents become hubs, and which indices become discovery gateways. This is also why Structure Engineering cannot be reduced to ``better prompts'' or ``more tools'': even if the View is clean (Layer 1), the system can still fail if it binds to the wrong capability, binds in the wrong order, or binds across incompatible assumptions. Structure Engineering therefore treats binding as a first-class architectural event and asks how to make binding outcomes predictable and scalable.

\textbf{Challenge: Self-Organization Entropy \& Binding Failures}
We define \textbf{Self-Organization Entropy} as the uncertainty of binding failures while the system moves from its \emph{current} binding topology toward a topology that would be most favorable for the current task. The core issue is not the gap itself; it is the uncertainty in whether the system can reliably reach the favorable topology under weak constraints. As the number of artifacts and agents scales, the combinatorial candidate space expands while prior constraints remain weak, which makes the formation of a useful topology increasingly unstable~\cite{liu2023agentbench,xia2024eddops}. In practice, this often manifests as:
\begin{itemize}
    \item \textbf{Binding Miss:} The system fails to find a suitable artifact or agent to bind to, so progress stalls, loops, or falls back to low-confidence guesses~\cite{asai2023self,liu2023agentbench}.
    \item \textbf{Binding Wrong:} The system binds to an unsuitable artifact or agent (e.g., an incompatible capability, an irrelevant index, or a peer that cannot deliver the required output), leading to incorrect execution paths or non-convergent coordination~\cite{patil2023gorilla,qin2023toolllm,liu2023agentbench}.
    \item \textbf{Binding Too Much:} The system binds to too many artifacts and agents. This over-exposes candidates and their metadata in the View, causing context pressure and pollution; it can also create an overly dense topology that increases coordination overhead and reduces task efficiency~\cite{packer2023memgpt,rath2026agentdrift}.
\end{itemize}
These failures share a common cause: the system is trying to construct a task-appropriate topology under weak constraints. When routing is under-specified, it compensates by guessing; when multiple partially relevant candidates exist, it oscillates; and when bindings are made without clear contracts, the resulting topology becomes brittle and hard to debug.

\subsection{Design Principles}
LSS systems are intelligent enough to self-organize and adapt their collaboration topology to different tasks, but this creates an engineering dilemma: binding cannot be left completely free, and it also cannot be fully hard-coded as in traditional software. Structure Engineering therefore aims to keep control on the ``edge of chaos'' through semantic design. To tame self-organization entropy without over-constraining semantic adaptability, we propose three governance principles.

\textbf{Task-Scoped Modularity.}
Artifacts and agents that participate in the same task often exhibit stronger cohesion: within-task bindings are typically more frequent and reuse-oriented, while cross-task bindings should be mediated by explicit semantic gates~\cite{hong2023metagpt,qian2024chatdev,wu2023autogen}. A task cluster can start from a small seed set of agents/artifacts and expand along high-confidence, task-oriented edges. Different task clusters can still be connected through semantic gates, which helps reduce cross-task misbinding and prevents unbounded search from overwhelming the View.

\textbf{Binding Provenance.}
Provenance can be recorded for binding events (routing decisions, retrieval hits, tool calls, and inheritance spawns). Provenance turns ``why did it bind to that?'' from a semantic black box into a traceable explanation: which artifact introduced a dependency, what evidence justified it, and how downstream steps relied on it. This is essential for debugging misbindings for both human users and higher-level agents~\cite{ojewale2026audittrailsaccountabilitylarge,souza2025llmagentsinteractiveworkflow}.

\textbf{Structure as Ability.}
Delivering a structure to an agent is not delivering a single, precise answer; it is empowering the agent to find, compose, and validate answers under constraints. For example, \textbf{delivering a file is a direct information delivery} that requires unusually strong mutual understanding of what ``the right file'' is. \textbf{Delivering a file-system organization (e.g. a tree or a graph) instead delivers an ability:} the receiver can locate the right file by probing the structure with its own task needs. The same applies to multi-agent collaboration. Delivering a team structure---roles, connections, etc.---gives an agent the ability to coordinate and reuse peers without re-negotiating bindings from scratch. In this sense, indices, team specifications, and inheritance designs are structures that compress future binding uncertainty into more reliable, callable affordances. Even a skill can be abstracted as a structure that shapes tool invocation and agent behavior. 

This concept is abstract, so let us illustrate it with a concrete example. Consider an agent tasked with debugging a massive codebase. Giving the agent the full text of 5,000 source files (or even just the file names) leads to context collapse. However, delivering a hierarchical \texttt{index.md} (a Structure) that maps features to files does more than provide a specific file. It equips the agent with the capability to systematically navigate, probe, and isolate the desired files in this file system. The structure acts as an executable map that the LLM's reasoning engine can ``run,'' thereby providing an ability rather than a single answer.

\begin{center}
\begin{tcolorbox}[colback=gray!10, colframe=black!50, arc=2mm, boxrule=1pt, width=0.95\linewidth]
\textbf{\textit{Comparison}:}
\begin{itemize}
    \item \emph{\textbf{Traditional software:} Structure constrains Information. Information processing defines capability.}
    \item \emph{\textbf{LSS:} Information Emerges Structure. Structure extends capability.}
\end{itemize}
\end{tcolorbox}
\end{center}

In the patterns below, we call the structure among artifacts an \emph{Index}. The structures that relate artifacts/temporally evolving information and agents can be defined by the previous layer's \emph{Semantic Lens} and, in this layer, by the \emph{Semantic Router}. For agent--agent structure, we distinguish two families: (i) communication-based cooperation topologies, which we call a \emph{Team}; and (ii) \emph{agentic inheritance}, which structures how View and Context are composed across derived agents.

\subsection{Design Patterns}

Following the above principles, we can give a set of structural-level design patterns.

\textbf{Semantic Router.}
\emph{Intent.} Route information to the right agent under semantic constraints.
\emph{Mechanism.} Given a piece of information---for example an Intent (what needs to be done next) or an output (what has just been produced)---a Semantic Router forwards it to the most suitable agent. This is different from \emph{Semantic Lens}: Lens selects proper information (Artifacts) to place into a given agent's View; Router decides which agent should receive a given message. The router can also record binding provenance for its routing decisions so that downstream agents can understand why a route was chosen. More subjectively, the Semantic Lens constructs an agent's View and thus acts like an \emph{agent-class generator}, whereas the Router primarily determines the inputs to an agent class, effectively instantiating the class.

\textbf{Index Generator.}
\emph{Intent.} Enable an agent to discover relevant Artifacts.
\emph{Mechanism.} An Index Generator produces an indexing system for the Artifacts relevant to the current task, such as a tree, a graph, or a semantic linked list. Since Artifacts are files, the Index can be viewed as a generated, task-specific file system. The index is a navigational structure that makes intentional discovery easier and more reliable than free-form searching. The index can be recorded and exchanged as \texttt{index.md}. A particularly useful form is a file-centered pointer graph that links semantically related files. Given a focal file, the Index Generator can produce a neighborhood structure centered on it, where each edge explains how the linked file relates to the focal one.

\begin{center}
\begin{tcolorbox}[colback=gray!10, colframe=black!50, arc=2mm, boxrule=1pt, width=0.95\linewidth]
\textbf{\textit{Comparison:}}
\begin{itemize}
    \item \emph{\textbf{Traditional software:} The file system is a rigid, objective hierarchy manipulated by explicit paths.}
    \item \emph{\textbf{LSS:} The file system (Index) is a fluid, subjective structure generated on-the-fly by semantic Intent.}
\end{itemize}
\end{tcolorbox}
\end{center}

It is also interesting that Semantic Lens can be viewed as a kind of Index. Unlike an explicit tree or graph structure, it is an intelligent index whose structure is largely a black box. Therefore, if an agent can generate a ``Semantic Lens'', that generator can also be seen as an Index Generator. This example highlights a key difference from traditional software design: in LSS, a role or an abstraction can shift with the designer's/agents' perspective. We call this subjectivity of design viewpoints the \emph{polymorphism of abstractions}; together with their composability (e.g., packaging skill-defined services as MCP tools), it reflects the mindset shift required when moving beyond traditional software thinking. These fluid concepts can further be generated, evolved, and nested at runtime, which gives LSS a very flexible and inclusive design space.

\textbf{Team Generator.}
\emph{Intent.} Enable an agent to collaborate with other agents.
\emph{Mechanism.} A Team Generator produces a cooperation structure among agents for a given task: roles, who talks to whom, and what each agent is responsible for. The team specification can be written and exchanged as \texttt{team.md}. With an explicit team, routing becomes a structural operation rather than an ad-hoc choice. A team can include Mediators and Routers to keep collaboration well-structured. And similarly, a Router can be viewed as an intelligent team structure.

\begin{figure}[h]
  \centering
  \IfFileExists{Figure/Figure_fluid_team.png}{\includegraphics[width=0.8\linewidth]{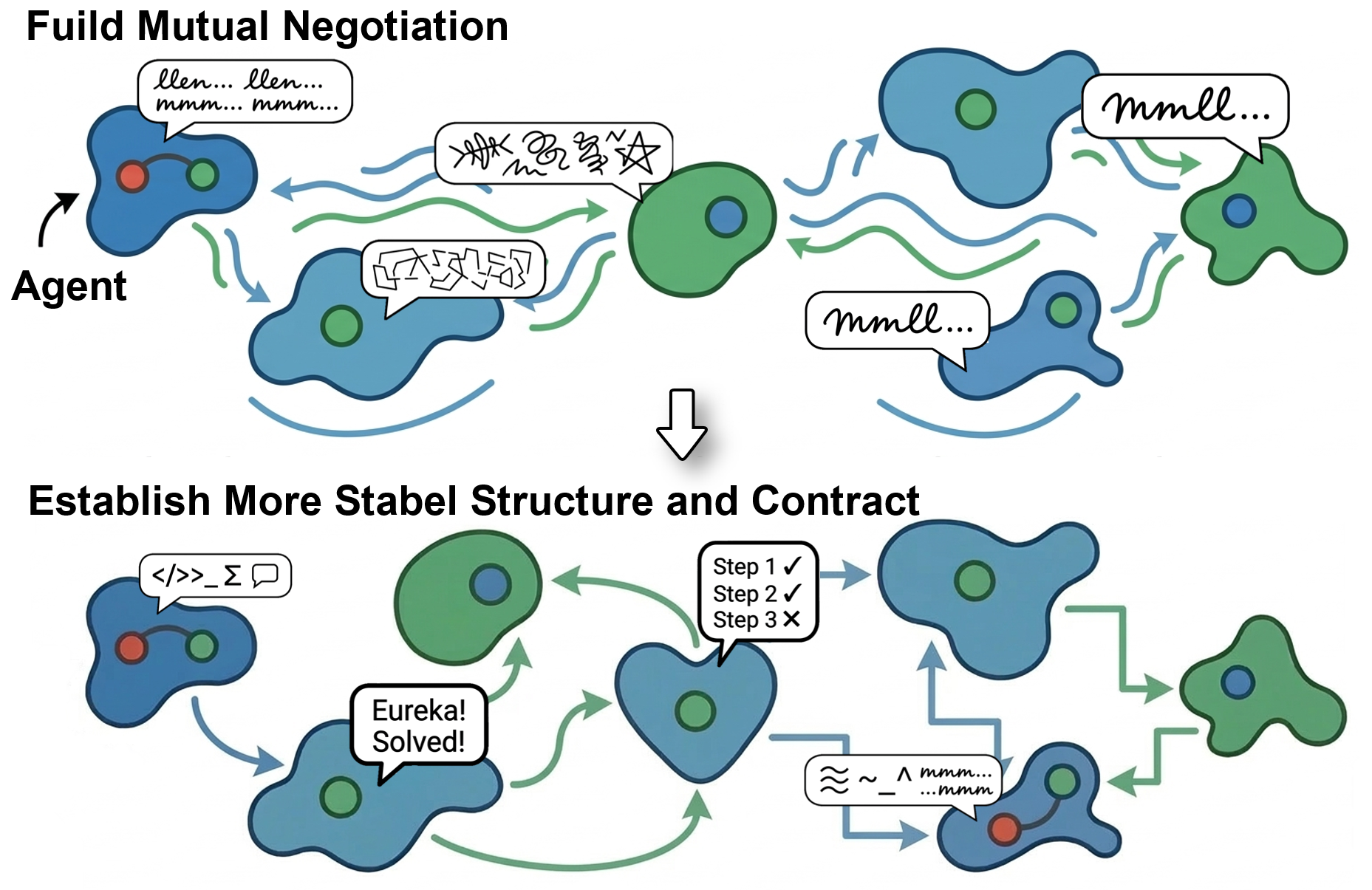}}{\fbox{\parbox{0.7\linewidth}{\centering Placeholder: \texttt{Figure/Figure\_fluid_team.png}}}}
  \caption{Establishing a more stable structure and contract through flexible multi-agent negotiation.}
  \label{fig:fluid-team}
\end{figure}

\begin{center}
\begin{tcolorbox}[colback=gray!10, colframe=black!50, arc=2mm, boxrule=1pt, width=0.95\linewidth]
\textbf{\textit{Comparison:}}
\begin{itemize}
\item \emph{\textbf{Traditional software:} Components integrate through rigid, syntactic contracts (APIs).}
\item \emph{\textbf{LSS:} Agents integrate through fluid, semantic negotiation (Natural Language Contracts) (Figure~\ref{fig:fluid-team}).}
\end{itemize}
\end{tcolorbox}
\end{center}

\textbf{After-task team.} Sometimes the team can be created \emph{after} a task: agents first self-organize to complete the task, and then the successful collaboration experience is summarized into \texttt{team.md} as a persistent team structure~\cite{xia2024eddops}. This can be viewed as an evolution mechanism discussed in the next section. A \texttt{team.md} itself can also serve as a task-completion playbook for future agents.

\begin{figure}[h]
  \centering
  \IfFileExists{Figure/Figure_speculative_branching_selection.png}{\includegraphics[width=0.9\linewidth]{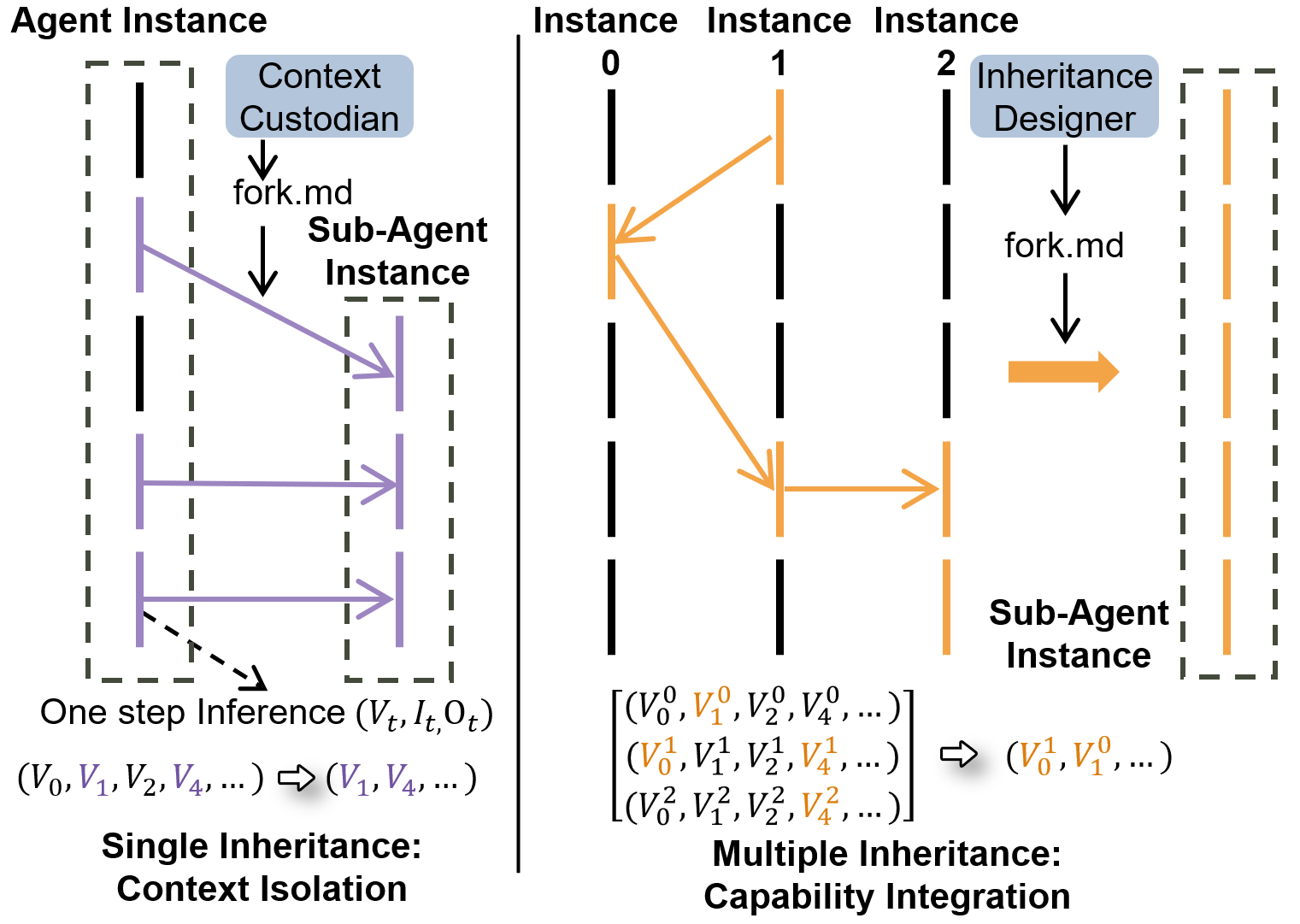}}{\fbox{\parbox{0.95\linewidth}{\centering Placeholder: \texttt{Figure/Figure\_speculative\_branching\_selection.png}}}}
  \caption{Inheritance in LSS: single inheritance isolates a child agent's context under a Context Curator; multiple inheritance selects and composes trajectory fragments across instances under an Inheritance Generator to integrate capabilities and memories.}
  \label{fig:semantic-inheritance-isolation-integration}
\end{figure}

\textbf{Inheritance Generator.}
\emph{Intent.} Derive a task-fit, context-clean agent by inheriting from existing agent instances.
\emph{Mechanism.} For a specific step, the Inheritance Generator derives a child agent from one or more existing agent instances, inheriting only the minimal required trajectory and constraints while keeping the child's View clean and avoiding pollution of the parent. This resembles a \texttt{fork} operation in modern OSes. The Inheritance Generator can also govern the lifecycle of child agents by specifying when to fork and by applying End Criteria. When there is only one parent, the Inheritance Generator reduces to the Context Curator introduced in the previous section. Therefore, in this section, we emphasize \emph{multiple inheritance}: the primary responsibility is not merely cleaning/compressing a parent's context, but integrating and migrating capabilities composed from multiple agent instances to produce an agent that is better fit for a specific task. The inheritance structure can be recorded in \texttt{fork.md}. Figure~\ref{fig:semantic-inheritance-isolation-integration} contrasts single-inheritance context isolation with multiple-inheritance capability integration.

While we intentionally borrow terminology from OOP (such as Classes, Inheritance), we emphasize that LSS reinterprets these concepts from a syntactic space to a semantic space. In classical OOP, inheritance is a static, compile-time relationship defining structural reuse. In LSS, Semantic Inheritance is a dynamic, runtime event where a child agent is instantiated by inheriting the precise contextual trajectory (the semantic state), allowing execution to branch safely without polluting the main reasoning thread. (This runtime derivation somewhat resembles prototype-based programming in, e.g., JavaScript.) By mapping fluid AI behaviors to established software engineering behaviors, we provide a cognitive-bridge from traditional software engineering for governing prompt-driven execution.

Using Inheritance Generator, a parent agent can fork several isolated child branches to explore different binding hypotheses (e.g., alternative retrieval paths, routing choices, or binding orders) without polluting the main execution. Each branch runs under a small contract (scope, budget, allowed side effects, and expected return). A selector then compares outcomes and merges only distilled, verifiable deltas back into the parent---for example, a validated index entry, a corrected routing rule, or a reliable invocation recipe---discarding the rest.

\textbf{Supply Chain.}
\emph{Intent.} Make multi-hop binding explanations traceable and stable.
\emph{Mechanism.} Provenance is a property of each binding event. By linking together the provenance of multiple bindings, the system forms a \emph{supply chain} across artifacts and agents: which routing decision led to which retrieval, which retrieval led to which invocation, and which invocation produced which downstream dependency---and, crucially, why each hop was justified. Such chains can be maintained at runtime and viewed as a semantic, distributed logging mechanism or a call stack~\cite{ojewale2026audittrailsaccountabilitylarge,souza2025llmagentsinteractiveworkflow}.

\textbf{Facade \& Filter.}
\emph{Intent.} Encapsulate a clean semantic gate for complex subsystems.
\emph{Mechanism.} Similar in spirit to the OOP Facade Pattern, the system exposes a simplified external interface for a complex internal collaboration subgraph, hiding internal negotiations and intermediate artifacts. In practice, the core of OpenClaw can be viewed as such a Facade with agent Router---a local gateway~\cite{openclaw2026}. Unlike a purely syntactic API wrapper, the facade is semantic: it can filter and reorganize inbound and outbound information (e.g., redact sensitive details, enforce policy constraints, or normalize outputs into a stable schema). Metaphorically, a Facade can act like a ``Maxwell's demon'' that reduces runtime entropy in the system by preventing unnecessary or harmful information passing and reducing failures caused by mismatched bindings.

It is worth noting that dynamic or static skill creators are mechanisms widely supported by current agent frameworks; they can also be seen as a design pattern for constructing structure among tool calls and agent behaviors.

\section{Evolution Engineering: Taming Evolutionary Entropy}\label{sec:layer3}

The third layer marks the transition from designing \emph{program execution} to governing \emph{endogenous program evolution}. An LSS does not merely run; it continuously rewrites the Artifacts that make it functional. More concretely, an LSS \emph{is} a collection of Artifact files. Agent capabilities, team organization, tools and data, prompts, skills, indices, and routing policies are all files that actively determine the system's behavioral boundary. They shape what can be bound, which Views are constructible, and which actions are reachable. Evolution Engineering studies this reality directly. If the system can change these persistent files from inside its own execution loop, then ``running the program'' and ``changing the program'' collapse into one continuous process, and \emph{using the system becomes part of developing it}.

\begin{center}
\begin{tcolorbox}[colback=gray!10, colframe=black!50, arc=2mm, boxrule=1pt, width=0.95\linewidth]
\textbf{\textit{Comparison:}}
\begin{itemize}
    \item \emph{\textbf{Traditional software:} Developing for usage.}
    \item \emph{\textbf{LSS:} Usage as part of development.}
\end{itemize}
\end{tcolorbox}
\end{center}

In this view, \textbf{Endogenous Evolution} is the long-term self-development of the system by modifying its own Artifacts. Because these changes happen inside the execution loop, they create a feedback cycle. The same View-defined inference that produced a behavior can also justify a patch that changes the Artifact behind that behavior, and the patched Artifact then shapes future Views that justify future patches. This is less like editing a static codebase with a clear split between development and deployment and more like online incremental learning on a moving target, where biases can be amplified and friction reduced through repeated reuse~\cite{han2025atp,rath2026agentdrift}.

Note that Evolution Engineering in this Section mainly focuses on the \emph{system scale}: how artifacts are created, revised, retained, and retired. Improvements to the base model's capability are outside the engineering boundary here. It also distinguishes \textbf{cross-session evolution} from \textbf{in-context adaptation}. A single agent may improvise tactics within one run to finish a task. Evolution Engineering is about what the system \emph{keeps} across runs: stable, structural improvements and capability deposits that change what future runs can do. Therefore, the evolution engineering goal is to make the system \emph{better under repeated use}:
\begin{itemize}
    \item \textbf{More adaptive:} as it completes tasks and interacts with users, it becomes increasingly aligned with user preferences and working conventions, reducing context entropy by stabilizing what information is relevant and how it is presented.
    \item \textbf{More coherent:} internal structure becomes easier to navigate and compose. Early versions often carry visible hand-designed constraints and internal mismatches across the LSS. Two agents may lack an efficient communication protocol, interfaces may be inconsistent, and Artifacts may be formatted or scoped in incompatible ways. Over time, the Artifact set can also accumulate stale, redundant, or conflicting items. The system then behaves like a machine whose gears do not mesh: it turns with friction. Evolution should make these gears engage smoothly, reducing structural entropy.
    \item \textbf{More capable:} new, reusable abilities emerge as Artifacts and are consolidated into stable services.
    \item \textbf{Bounded blast radius:} failures are increasingly confined to small, local mistakes rather than spreading into system-wide drift.
\end{itemize}

\textbf{Challenge: Evolutionary Entropy (Knowledge Rot \& Long-term Drift).}
We define \textbf{Evolutionary Entropy} as the long-term uncertainty and unpredictable system drift induced by self-modification. After many edits, it becomes hard to tell what truly helped and what else was unintentionally affected. In practice, it often manifests as:
\begin{itemize}
    \item \textbf{Too lazy:} evolution is too slow. Valuable information and working methods discovered in tasks are not retained, so the system cannot accumulate useful capability.
    \item \textbf{Too active:} evolution is too frequent. Behavior becomes unstable, changes oscillate, and the Artifact store fills with debris and inconsistencies, accelerating knowledge rot~\cite{rath2026agentdrift}.
    \item \textbf{Goal misalignment:} evolution does not move toward the human or task intent. Over repeated self-evolution cycles, locally rewarded patches can compound into systematic deviation from human intent as a post-deployment failure~\cite{han2025atp}.
\end{itemize}
What is hard for evolution engineering is that a patch can help on the current task while making the overall system worse. Evolution needs to learn from short-term success, but it also needs a long-term direction. This requires the evolving system to model its own operation and objectives, and to apply changes in a planned, deliberate way.

\subsection{Design Principles}
\begin{figure}[h]
  \centering
  \IfFileExists{Figure/Figure_ephemeral_vs_persistent.png}{\includegraphics[width=1\linewidth]{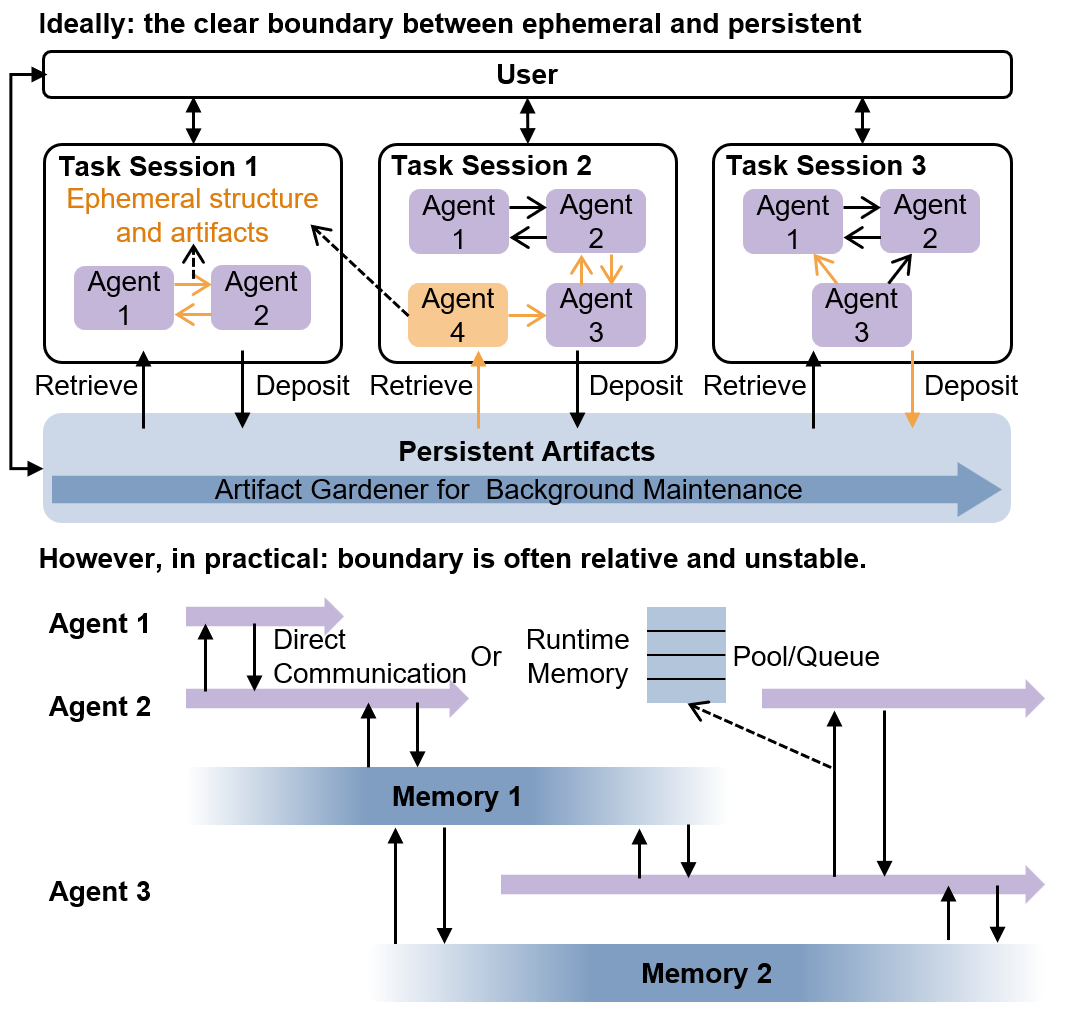}}{\fbox{\parbox{0.8\linewidth}{\centering Placeholder: \texttt{Figure/Figure\_ephemeral\_vs\_persistent.png}}}}
  \caption{The evolution engineering should accept a relative and unstable ephemeral/persistent Boundary.}
  \label{fig:ephemeral-vs-persistent}
\end{figure}

\textbf{Plan the Ephemeral/Persistent Boundary.}
Since an LSS is a collection of Artifact files, Evolution Engineering is mainly about planning the boundary between what is temporary for a task and what becomes part of the persistent Artifact set. At its simplest, the LSS maintains a global persistent Artifact pool, and the Evolver deposits important, effective ephemeral structure and Artifacts from each task session into this pool; an example of such a structure is OpenClaw’s two-level memory~\cite{openclaw2026}. However, ``ephemeral vs.\ persistent'' is not a binary property but a time-scale-relative spectrum (Figure~\ref{fig:ephemeral-vs-persistent}): a memory item may be persistent over a day yet ephemeral relative to a week-long redesign; a routing rule may be stable for a project phase but later become a transient cache after a project finishes. Across multi-scale agent lifecycles and multi-layer memories, different layers can have different levels of evolutionary activity.

\textbf{Evolution as Learning.}
Evolution is a learning process that needs environmental feedback. Useful feedback signals include:
\begin{itemize}
    \item \textbf{Objective verification:} whether the system outputs satisfy tests, constraints, or user-specified checks.
    \item \textbf{Human feedback:} corrections, suggestions, accept/reject behavior, rollbacks, and edit traces as ground truth for alignment.
    \item \textbf{Self-reflection:} internal evaluation of process and outcome, e.g., a reviewer/critic agent iteratively provides feedback to a worker agent for refinement~\cite{madaan2023selfrefine}.
\end{itemize}

\textbf{Hebb's Rule.}
``fires together, wires together'' is a local, unsupervised learning rule inspired by synaptic plasticity~\cite{hebb1949organization}. For LSS, the analogous engineering principle is to reinforce structural pathways that are repeatedly used across tasks. When a binding path, skill, index, or collaboration topology is repeatedly validated, the system can stabilize it by promoting it into a persistent artifact or wrapper in an explicit contract (e.g., an MCP or A2A contract). Conversely, changes correlated with regressions can be down-weighted or pushed back into ephemeral scope.

\textbf{Entropy-aware Evolving.}
Evolution should stay aware of the three entropies introduced in this paper. One practical requirement here is \textbf{history residues}: evolution should not be purely overwriting. The system can retain historical slices and their rationales so that it can roll back, and past trajectories act as a prior that smooths abrupt oscillations during exploration.

\textbf{Design the Structure of Evolving.}
Finally, the engineer's job shifts from hand-authoring rules for how the structure should evolve to building a structure that makes evolution possible and safe. This means trusting the LSS's own intelligence to propose improvements leaning toward \emph{endogenous and spontaneous evolution}, while providing \emph{meta-structure} (incentives, exploration space, selection pressure) that keeps evolution inside a desired scope.

\begin{center}
\begin{tcolorbox}[colback=gray!10, colframe=black!50, arc=2mm, boxrule=1pt, width=0.95\linewidth]
\textbf{\textit{Comparison:}}
\begin{itemize}
    \item \emph{\textbf{Traditional software:} Design the evolution rules of the structure.}
    \item \emph{\textbf{LSS:} Design the structure of evolution.}
\end{itemize}
\end{tcolorbox}
\end{center}

\subsection{Design Patterns}
Following the above principles, we can give a set of system evolution design patterns.

\textbf{Sandbox Mode.}
\emph{Intent.} Bound blast radius by default while still enabling fast exploration.
\emph{Mechanism.} When the system explores changes to persistent artifacts, it creates one or more sandbox environments that run in isolated scopes. It works on sandbox copies of the relevant artifacts and write targets. Comparable tasks (or replay tasks) are executed to evaluate each candidate, including A/B-style comparisons when feasible. Because the ephemeral/persistent boundary is relative, nested sandboxes could be introduced. A dedicated agent can decide when to spawn sandboxes and let the Evolver merge their results. Only changes that remain safe and consistently helpful are merged back into the persistent artifact set; others are discarded.

\textbf{Evolver.}
\emph{Intent.} Improve the system by observing and revising its artifact set over time.
\emph{Mechanism.} The Evolver continuously monitors task completion, interaction traces, and user feedback, then proposes concrete edits to the artifact set. Each proposal is packaged as an explicit, reviewable patch with a stated hypothesis (what behavior it should change) and a rollback chain. The Evolver gates merges through verification: it may replay representative tasks, run tests and static checks, and perform sandboxed A/B comparisons before promoting changes into the persistent artifact set. The behavior of Evolver can be specified in \texttt{evolve.md}.

As a concrete example, an Evolver can implement an agent-level genetic algorithm: starting from a baseline agent instance and its traces, it generates a population of derived candidate instances via counterfactual edits and controlled ``mutations'' (e.g., perturbing a binding choice, swapping a tool return, or injecting/removing a small context fragment) while keeping external writes isolated. Each candidate is evaluated by replaying tasks or running representative checks in sandbox mode, producing fitness signals. The Evolver then applies selection pressure by retaining the best-performing instances (and optionally recombining their compatible improvements) and discarding the rest, thereby performing an explicit natural-selection loop over agent instances. Finally, the Evolver distills valuable lessons and memories from the top-performing agent instances, updates the artifacts accordingly, and thus drives system evolution. This entire process can be repeated during idle periods by continuously replaying previously completed tasks.

\begin{center}
\begin{tcolorbox}[colback=gray!10, colframe=black!50, arc=2mm, boxrule=1pt, width=0.95\linewidth]
\textbf{\textit{Comparison:}}
\begin{itemize}
\item \emph{\textbf{Traditional software:} An exception breaks the control flow.}
\item \emph{\textbf{LSS:} An exception is appended to the context flow to generate the next action.}
\end{itemize}
\end{tcolorbox}
\end{center}

\textbf{Semantic Palimpsest.}
\emph{Intent.} A new file type that carries continuous semantic history residues.
\emph{Mechanism.} Semantic Palimpsest is a file that stores more than the current version. It carries continuous semantic residues: how an Artifact evolved, which changes mattered for behavior, and why they were made. Even when the file displays a ``current'' version, the semantic shadow of its trajectory remains available and influential. Compared with version control systems like git, it can still support rollback, but it emphasizes semantic history over line-by-line diffs. Small edits that cause large behavior changes may be salient~\cite{errica2025did}, and the reasons behind changes stay attached to the Artifact so that the evolution path is explainable.

\textbf{Artifact Maintainer.}
\emph{Intent.} Counteract entropy accumulation in the Artifact store.
\emph{Mechanism.} An asynchronous maintainer scans for redundancy, outdated items, and fragmentation. It consolidates duplicates, marks stale items, detects conflicts, and retires structures that are rarely bound, keeping the Artifact space navigable and reducing the inventory cost of evolution. The Maintainer can also report warnings for structural incoherence.

\textbf{Artifact Tiering.}
\emph{Intent.} Make layered ``ephemeral vs.\ persistent'' a property of the file system.
\emph{Mechanism.} Memory and Artifacts are organized into explicit tiers with different retention and read/write policies:
\begin{itemize}
    \item \textbf{Hot tier:} fast-changing work products created, modified, and deleted frequently for active tasks.
    \item \textbf{Warm tier:} stable core skills, indices, and other long-term memories that are read often and written rarely.
    \item \textbf{Cold tier:} low-frequency, edge Artifacts that may dissipate or be forgotten.
\end{itemize}
Evolution is then not only content change but also \emph{tier migration}: promoting/demoting artifacts across tiers becomes an explicit operation that expresses lifecycle intent. This framing echoes a large body of research on MAS memory, where memory is treated as differentiated stores with different read/write policies, maintained by retrieval, summarization/reflection, consolidation, and forgetting to preserve long-horizon coherence under limited context~\cite{wang2024survey,liu2023lostmiddlelanguagemodels,lewis2020rag,park2023generativeagents,packer2023memgpt,zhong2023memorybank,wang2023voyager,shinn2023reflexion,xu2025amem}. Accordingly, we do not further delve into memory management here.

\textbf{Shared Interaction Space.}
\emph{Intent.} Establish a work-together space in which human and agent actions are both explicit, permissioned operations.
\emph{Mechanism.} A shared space supports (i) co-creation via joint editing of Artifacts, (ii) human specification via direct manipulation of objects (which acts as structured instruction), and (iii) agent learning by observing edit traces and acceptance behavior. This reduces ambiguity about intent and couples direct manipulation with automation and interactive learning from user feedback~\cite{christiano2017preferences,shinn2023reflexion}. Ideally, this shared space exposes humans to a stable file system, while the indices and internal working representations maintained by agents can remain more volatile and messy from a human perspective.

\begin{figure}[h]
  \centering
  \IfFileExists{Figure/Figure_evolution_patterns_triad.png}{\includegraphics[width=1\linewidth]{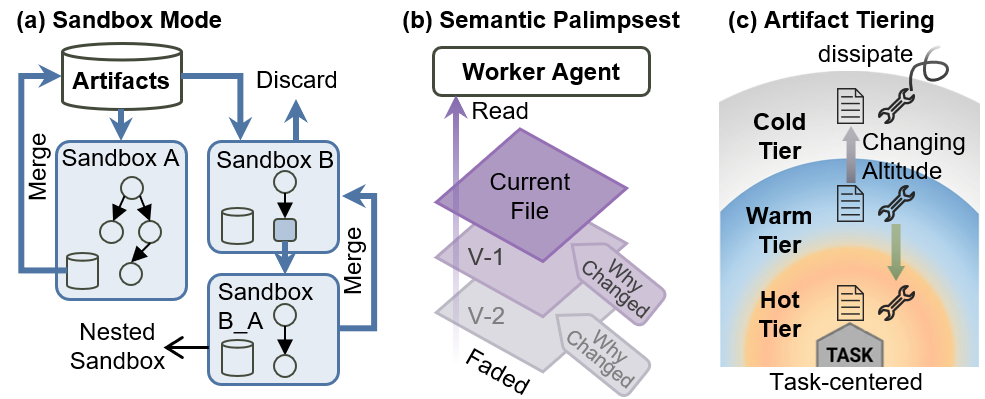}}{\fbox{\parbox{0.95\linewidth}{\centering Placeholder: \texttt{Figure/Figure\_evolution\_patterns\_triad.png}}}}
  \caption{Selected evolution patterns in Layer 3: Sandbox Mode evaluates changes before they enter the persistent artifact set, Semantic Palimpsest keeps changes traceable and rollback-capable, and Artifact Tiering manages artifact lifecycles by moving them across tiers with different read/write policies.}
  \label{fig:evolution-patterns}
\end{figure}

\section{Design Pattern Mapping}\label{sec:cohesion}

While LSS Design Patterns provide a top-down inspiration for system engineering, they function as logical guidelines rather than strict physical implementation. Unlike traditional OOP where a design pattern often corresponds directly to a class structure, an LSS pattern like "Router" does not necessarily mandate a dedicated "Router Agent." Instead, these patterns should be viewed as cognitive mechanisms or plugins. A single physical Agent might embody multiple logical patterns, or conversely, a single logical pattern might require a collaborative team of Agents to execute.

The mapping from logical design to physical Agents involves several strategies:
\begin{itemize}
    \item \textbf{Dedicated Agent:} Assigning a logical functional module to a specialized Agent.
    \item \textbf{Embedded Mechanism:} Implementing multiple logical modules as internal steps or built-in mechanisms within a single Agent.
    \item \textbf{Collaboration:} Utilizing a team of Agents to collaboratively implement a single logical module.
    \item \textbf{Derived Execution:} Deriving a new Agent from an existing one to execute a module. Suitable when sharing the parent's context is necessary, but the parent cannot afford additional context burden from the sub-process.
    \item \textbf{Post-Execution Derived:} Implementing the module directly within the current Agent's context, but deriving a fresh Agent after the execution to "forget" the temporary module implementation.
    \item \textbf{Self-Derived Loop:} An Agent uses a derivation loop to continuously derive its own ``next-hop instance'' to maintain execution continuity while refreshing its context at each iteration.
\end{itemize}

\begin{figure}[h]
  \centering
  \IfFileExists{Figure/Figure_semantic_lens_sequence.png}{\includegraphics[width=1\linewidth]{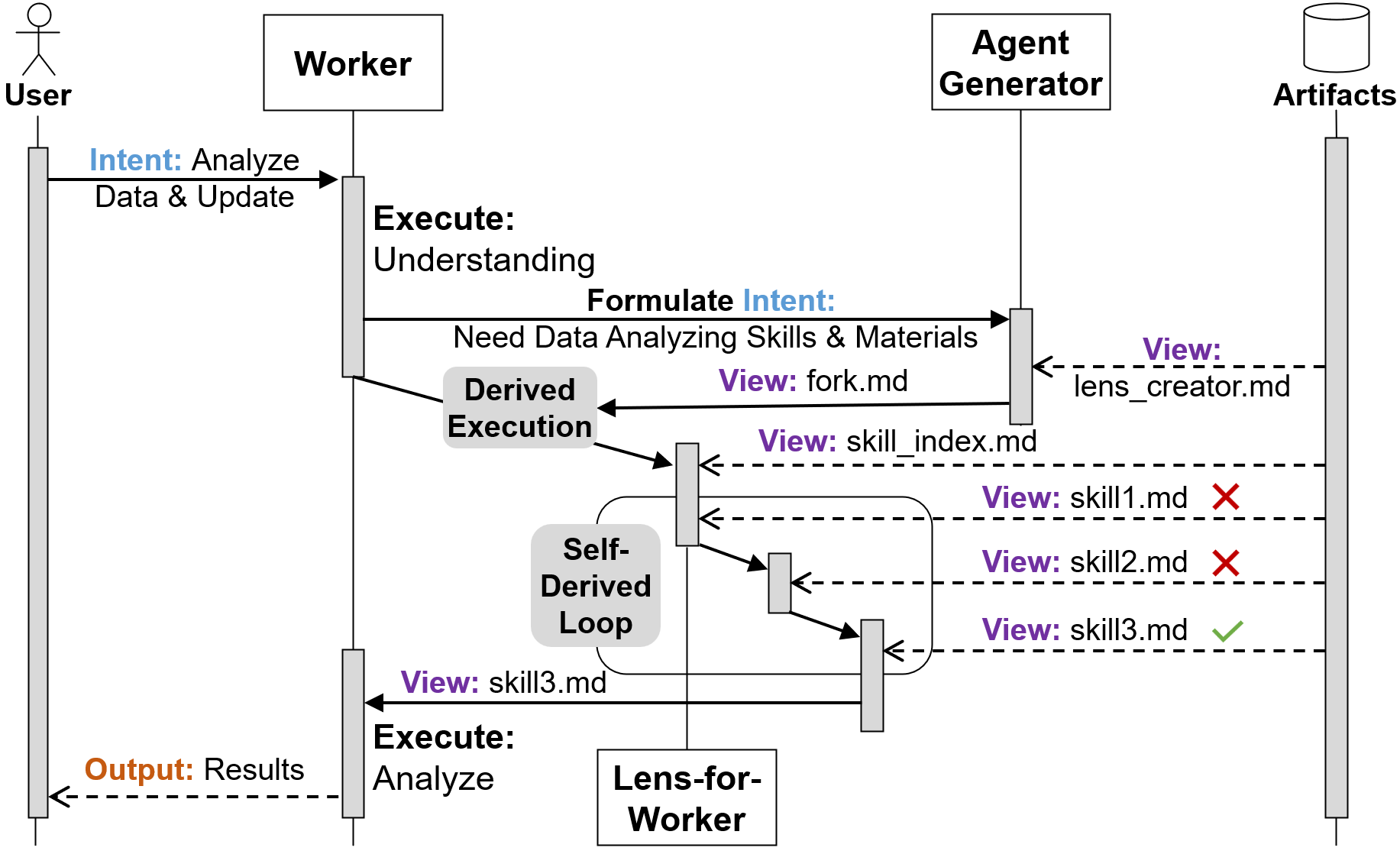}}{\fbox{\parbox{0.95\linewidth}{\centering Placeholder: \texttt{Figure/Figure\_semantic\_lens\_sequence.png}}}}
  \caption{Sequence diagram of the Semantic Lens pattern illustrating Derived Execution, Collaboration, and Self-Derived Loops. The Worker Agent initiates the process, and the Lens Agent recursively derives new instances to handle each artifact, keeping the Worker's context clean.}
  \label{fig:semantic-lens}
\end{figure}

Consider the ``Semantic Lens'' example introduced in Figure~\ref{fig:lss-sequence}. Figure~\ref{fig:semantic-lens} shows one possible mapping implementation. When a Worker Agent needs a specific skill artifact to complete a task, it can reuse its own context to derive a temporary ``Lens Agent,'' injecting it with the \textit{Intent} for the required file (\textit{Derived Execution}). In practice, this is often handled by a system-level Agent Generator, which invokes a Lens Agent skill based on the Worker's request (\textit{Collaboration}) and derives the Lens Agent's initial state from the Worker; exposing a Lens creator skill directly to the Worker is not a scalable design. The Lens Agent then iterates over a Skill Index, querying files one by one. After reading each file's content, it determines whether the content satisfies the Worker's needs. If so, it records the file path. Regardless of the match result, after processing a file the Lens Agent derives a fresh instance to clear the context occupied by the file it just read (\textit{Self-Derived Loop}). Finally, the aggregated paths are returned to the Worker, and the Lens Agent is terminated once its end criteria are met. The Worker remains uncontaminated by massive search noise.

\textbf{Design Principle: Semantic Cohesion Principle.}
One of the guiding principles for these architectural decisions is Semantic Cohesion, which replaces the Single Responsibility Principle (SRP)~\cite{martin2002agile} in OOP. While SRP separates classes based on functions, Semantic Cohesion groups capabilities based on their \textit{shared semantic information required for reasoning}. If two mechanisms rely on the same documents and memory, they should likely be merged to avoid the cost of mutual understanding among multiple agents. Conversely, if they require incompatible Views, or if we intentionally want to explore diverse Views, they are better split to prevent context pollution, even if they are functionally related. Additionally, the choice of mapping strategy should also consider the Agent's real-time context pressure.

\begin{center}
\begin{tcolorbox}[colback=gray!10, colframe=black!50, arc=2mm, boxrule=1pt, width=0.95\linewidth]
\textbf{\textit{Comparison:}}
\begin{itemize}
    \item \emph{\textbf{Traditional software:} Single Responsibility Principle for designing a class.}
    \item \emph{\textbf{LSS:} Semantic Cohesion Principle for mapping an agent.}
\end{itemize}
\end{tcolorbox}
\end{center}

\textbf{Design Pattern: Vibe Compiler.}
\emph{Intent.} Enable developers to define system architecture through loose, standardized descriptions rather than rigid code, and automate the physical agent mapping.
\emph{Mechanism.} We envision a \textit{Vibe Compiler} as a Meta Agent that takes a standardized, loose architectural description (akin to a UML class diagram) and automatically generates an LSS implementation satisfying SCP. It frees developers from being concerned about low-level agent-orchestration details. Empirically, such strongly semantic systems risk over-engineering: they need space for self-organization and evolution to maintain adaptability.

\textbf{Design Pattern: Runtime Pattern Shifter.}
\emph{Intent.} Dynamically adapt the system topology to real-time context pressure and task complexity.
\emph{Mechanism.} A \textit{Runtime Pattern Shifter} monitors execution metrics (e.g., context window usage, ambiguity) and triggers different mapping modes on the fly. This pattern can be integrated with the \textit{Team Generator} discussed in the previous section and also can enable evolution mechanisms.

\section{LSS Workflow Evaluation}\label{sec:workflows}

\subsection{Evaluation on RepoBench-R}

We use RepoBench-R as a retrieval-focused code-completion benchmark to validate two mechanisms: \emph{Semantic Lens} and \emph{Index Generator}. Specifically, we use the Python subset (\texttt{python\_cff}) with the \texttt{test\_easy} split. RepoBench-R~\cite{liu2023repobench} provides code-completion queries paired with a candidate pool of cross-file snippets drawn from the same repository, along with a gold snippet identifier indicating which retrieved context is most relevant for predicting the next line. This setup isolates the retrieval subproblem: given the local code context, select a small set of supporting snippets that would best enable a downstream model to continue the code.

We compare three retrieval configurations under the same model (DeepSeek API) and the same candidate budget. Each query starts from a lexical pre-selection pool, from which each variant outputs top-$K$ snippets with $K=5$. In addition, evidence length is explicitly bounded: candidate briefs for Lens ranking are truncated to 280 characters, and Worker-side snippet reads are truncated to 700 characters per selected item. (A) \textbf{Worker-only retrieval} asks a single Worker to both scan all candidates and select the top-$K$ snippets. (B) \textbf{Lens + Worker} splits responsibilities: a Lens selects the top-$K$ candidates using brief evidence, and the Worker then reads only the selected snippets. (C) \textbf{Lens + Index + Worker} further inserts an Index Generator that produces compact per-candidate index descriptions; the Lens makes its selection using these indexes rather than raw snippet text, after which the Worker still consumes only the selected snippets. Across variants we track total token cost, per-agent token usage, and average input-context tokens per agent.

\begin{figure}[h]
  \centering
  \includegraphics[width=1\linewidth]{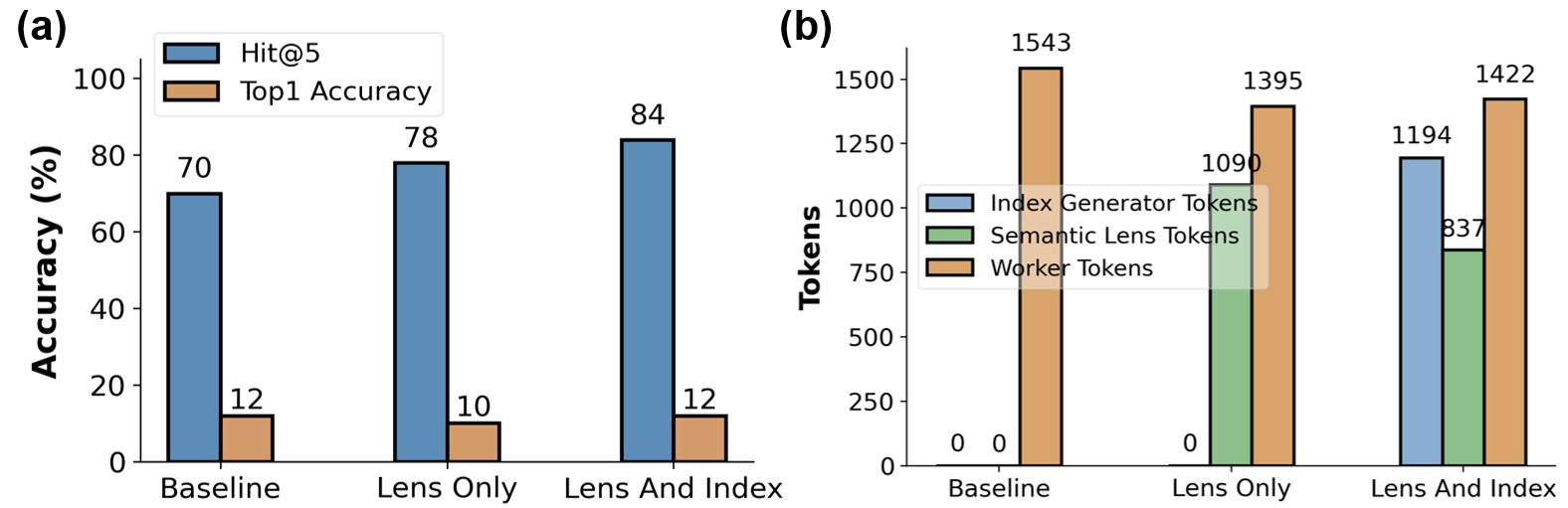}
  \caption{Evaluation results on RepoBench-R: (a) Retrieval success comparison, and (b) Average input-context tokens across variants.}
  \label{fig:lss-exp1-metrics}
\end{figure}

Retrieval quality is summarized in Figure~\ref{fig:lss-exp1-metrics}(a) using two complementary metrics. \textbf{Hit@5} measures recall: whether the gold snippet appears anywhere in the selected top-$5$ results. \textbf{Top1 Accuracy} measures precision at rank 1: whether the first selected candidate is exactly the gold snippet. On RepoBench-R, Lens+Worker improves Hit@5 over Worker-only (0.70 $\rightarrow$ 0.78), and Lens+Index yields the highest Hit@5 (0.84), indicating that indexing helps the Lens make more robust, recall-oriented selections. In contrast, Top1 Accuracy remains low and largely unchanged across variants (0.10--0.12), suggesting that while the mechanisms reliably broaden coverage of the correct evidence, ranking the single best snippet remains difficult and may require stronger reranking objectives or richer dependency signals.

To understand context pressure, we first examine average input-context tokens in Figure~\ref{fig:lss-exp1-metrics}(b). Introducing a Semantic Lens substantially reduces the Worker's average input (1543 $\rightarrow$ 1395) because the Worker no longer needs to ingest the full candidate pool. Adding the Index Generator keeps the Worker's average bounded at a similar level (1422) while shifting part of the selection burden from the Lens's raw-snippet evidence to compact index lines. The net effect matches the intended behavior: retrieval-related context pressure is redistributed away from the Worker toward specialized, bounded-scope agents.

\begin{figure}[h]
  \centering
  \includegraphics[width=0.9\linewidth]{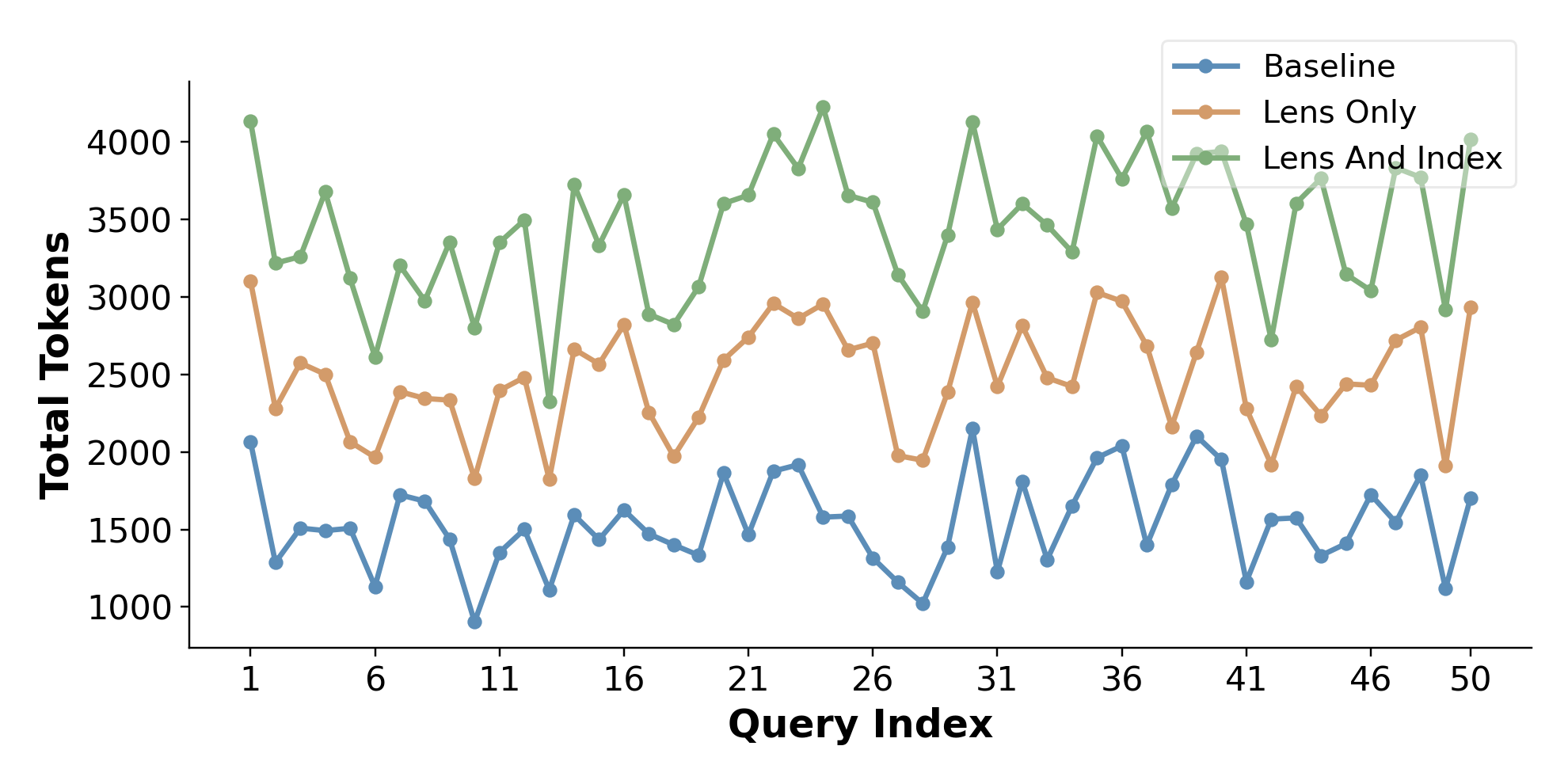}
  \caption{Per-query total token traces across three variants.}
  \label{fig:lss-exp1-total-scatter}
\end{figure}

Total token cost in Figure~\ref{fig:lss-exp1-total-scatter} increases for Lens-assisted variants because selection is externalized into additional agent calls. Although total token cost increases, the Index Generator cost is an amortizable overhead: if index packages are persisted and reused across many queries within the same repository (or the same stable candidate set), the one-time indexing cost can be spread over multiple Lens decisions, making the Lens+Index configuration attractive for repeated or long-horizon workflows.

\begin{figure}[h]
  \centering
  \includegraphics[width=0.9\linewidth]{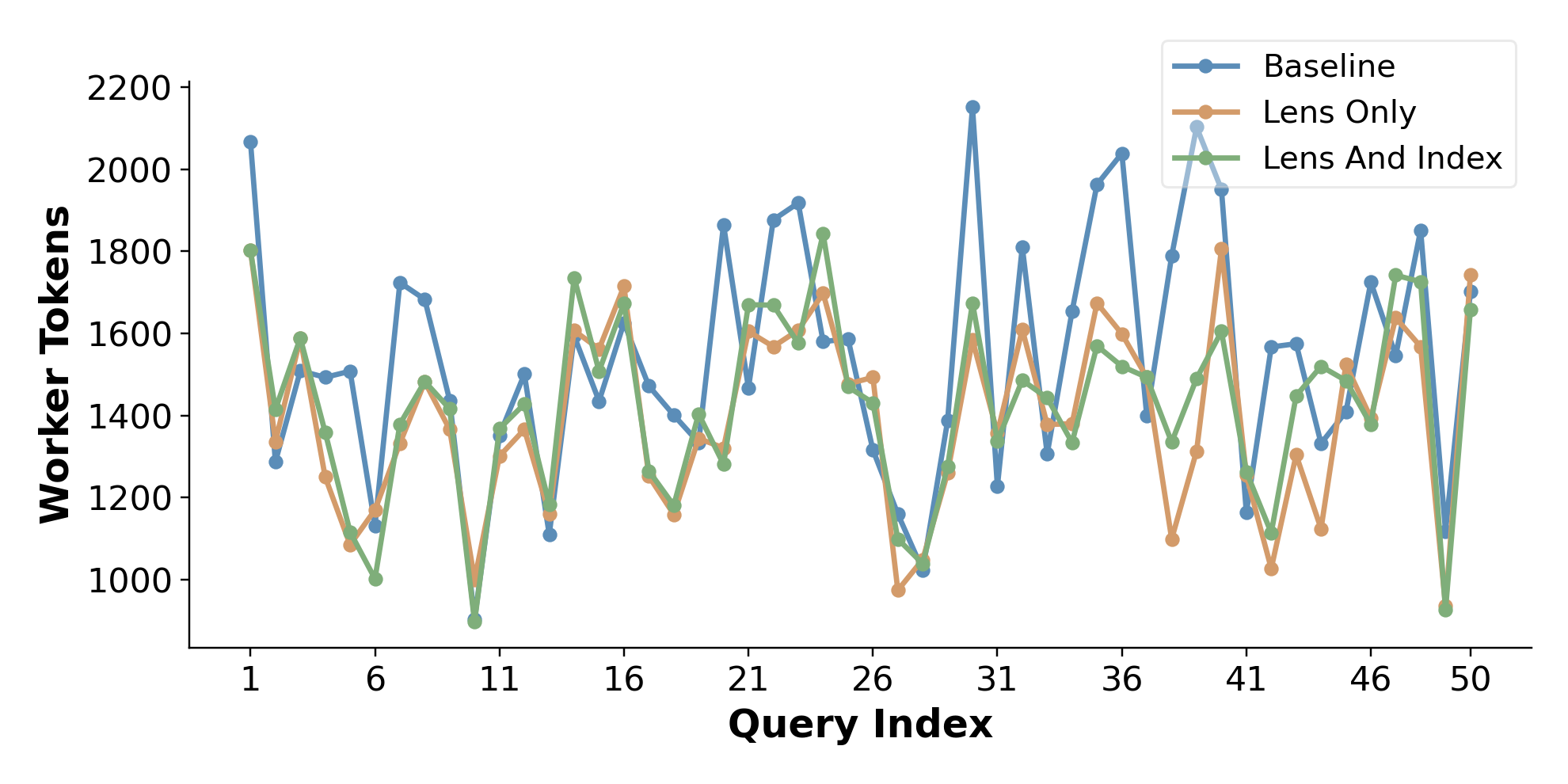}
  \caption{Per-query Worker token traces across three variants.}
  \label{fig:lss-exp1-worker-scatter}
\end{figure}

Worker-side token traces in Figure~\ref{fig:lss-exp1-worker-scatter} show that Lens-assisted routing often reduces Worker tokens and trims average context consumption, although the magnitude is query-dependent. Overall, RepoBench-R results demonstrate a quality--cost trade-off aligned with the LSS objective of controlled context routing: adding Semantic Lens and Index Generator improves recall-oriented retrieval (\emph{Hit@5}: 0.70 $\rightarrow$ 0.84) and bounds Worker context pressure, at the price of additional orchestration tokens that can be partially amortized via index reuse.

\subsection{Workflow on Comprehensive LSS Environment}

Beyond benchmark retrieval, we build an automated research environment to serve as a \emph{harness engineering} substrate for LSS workflows. At its core is a file-based project knowledge base that stores thousands of atomic entries---ideas, thoughts, references, experiment records, design decisions, and intermediate drafts---each with lightweight metadata for linking, status tracking, and incremental archival. The knowledge base grows continuously during day-to-day human--Agent collaboration: Agents can distill valuable artifacts from a human's daily outputs, and can also contribute new artifacts produced by literature review, code exploration, or experiment execution. Over time, this file-mediated knowledge base becomes a shared interaction space where both humans and Agents can read, write, and build upon the same persistent state.

\begin{figure}[h]
  \centering
  \includegraphics[width=0.9\linewidth]{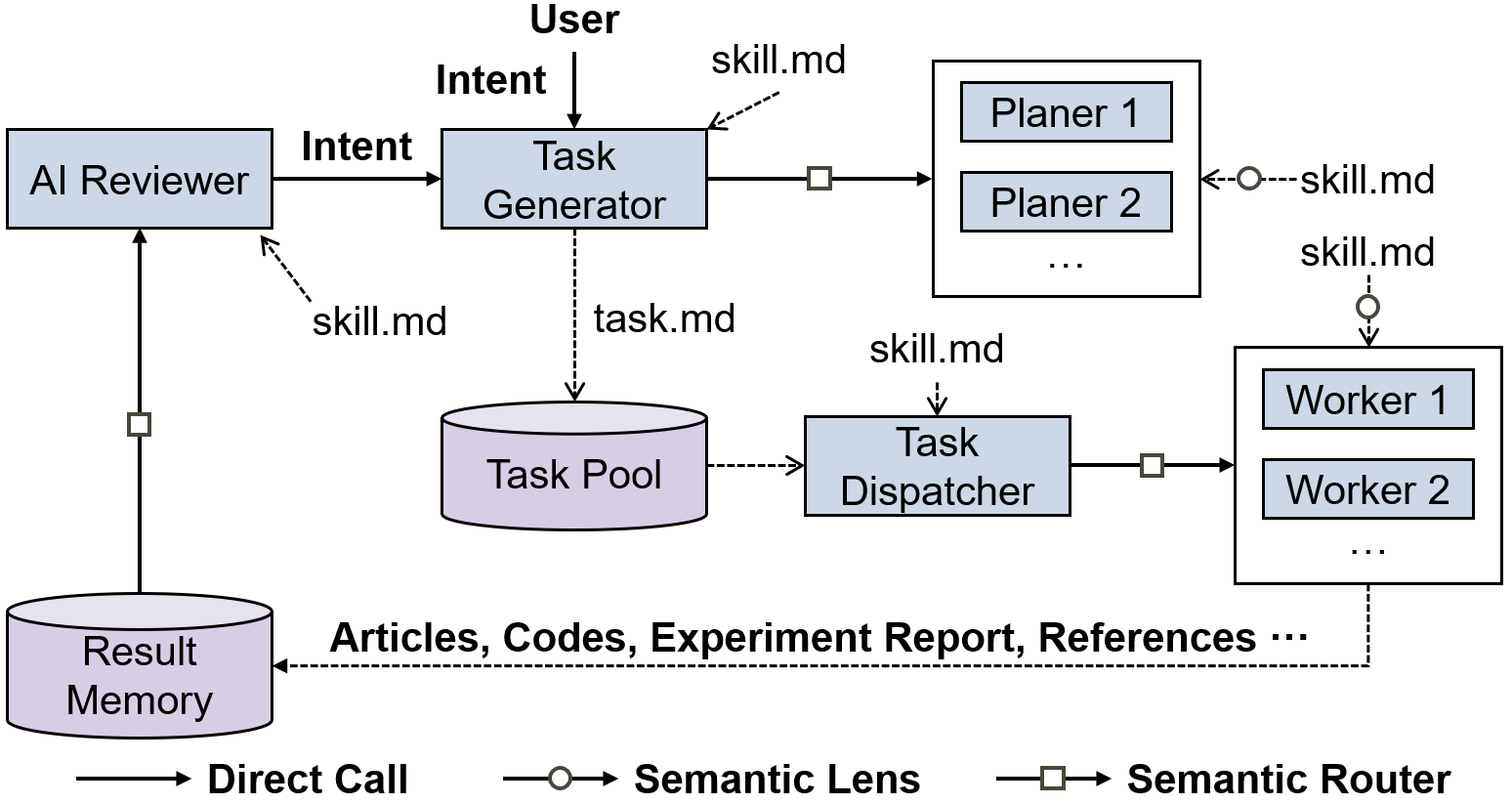}
  \caption{The agent class architecture of task distribution and completion workflow in the knowledge-base-centric environment.}
  \label{fig:lss-exp2-class}
\end{figure}

This knowledge base can act as both a productivity environment and an experiment environment for LSS: this paper itself was developed inside this environment, and this environment can be used for the evaluation of this paper. As a concrete example, we implement a file-mediated, auditable task distribution and completion workflow with persistent state. As shown in Figure~\ref{fig:lss-exp2-class}, a user request is first issued as an \emph{Intent}, which is then expanded into a set of concrete tasks written as structured \texttt{task.md} items into a shared \emph{Task Pool}. A semantic router selects appropriate planning/execution agents (e.g., Planner/Worker instances) conditioned on reusable capability specifications (\texttt{skill.md}); agents act over the same knowledge-base substrate (articles, code, experiment reports, references) and commit outcomes back to the filesystem as traceable artifacts. These agents proceed through the following steps:

\begin{enumerate}
    \item \textbf{Generate \& Dispatch:} An AI reviewer/controller transforms the user intent into tasks via a task generator, materializes them as \texttt{task.md} files in the Task Pool, and routes to find the suitable planners to facilitate the construction of \texttt{task.md}.
    \item \textbf{Execute \& Log:} A task dispatcher assigns (or workers claim) tasks from the pool; workers execute the actions, consult relevant skills/knowledge through semantic lenses, and append stepwise execution logs and outputs to the corresponding task files and the Result Memory.
    \item \textbf{Review \& Iterate:} The reviewer reads task summaries and accumulated results from the Result Memory, performs acceptance, and if requirements are not met, emits a new round of tasks; otherwise the workflow halts.
\end{enumerate}

We use such a workflow to assist us in completing a research process, but a full research cycle requires strong human interaction at key control points. For instance, manual review of \texttt{task.md}, fine-grained adjustment of article organization logic, and targeted investigations or experiments are often necessary. This makes the highly interactive process difficult to quantify. Furthermore, because it is challenging for agents to have a standardized metric for ``good research,'' fully automated research generation often struggles to meet our requirements. Therefore, after completing this paper, we tasked the aforementioned LSS workflow with reproducing the research process of this paper, but in a completely automated manner. We created an AI reviewer to evaluate the generated results of LSS by comparing them with the results of this paper and providing corresponding Intents to the Task Generator. Except for the AI Reviewer, no other LSS Agents had access to this paper. Since agents can always identify points for further improvement, we limited the number of \texttt{task.md} items produced by the Task Generator to a maximum of 10 per round. In total, 10 rounds of research activities were performed. Worker Agents were responsible for writing, proposing new ideas, conducting literature research (via Web Search tool), performing experimental verification of basic concepts, and generating figures. As illustrative figures cannot be generated directly, we configured the agents to produce prompts for Nano Banana.

\begin{figure}[h]
  \centering
  \includegraphics[width=0.95\linewidth]{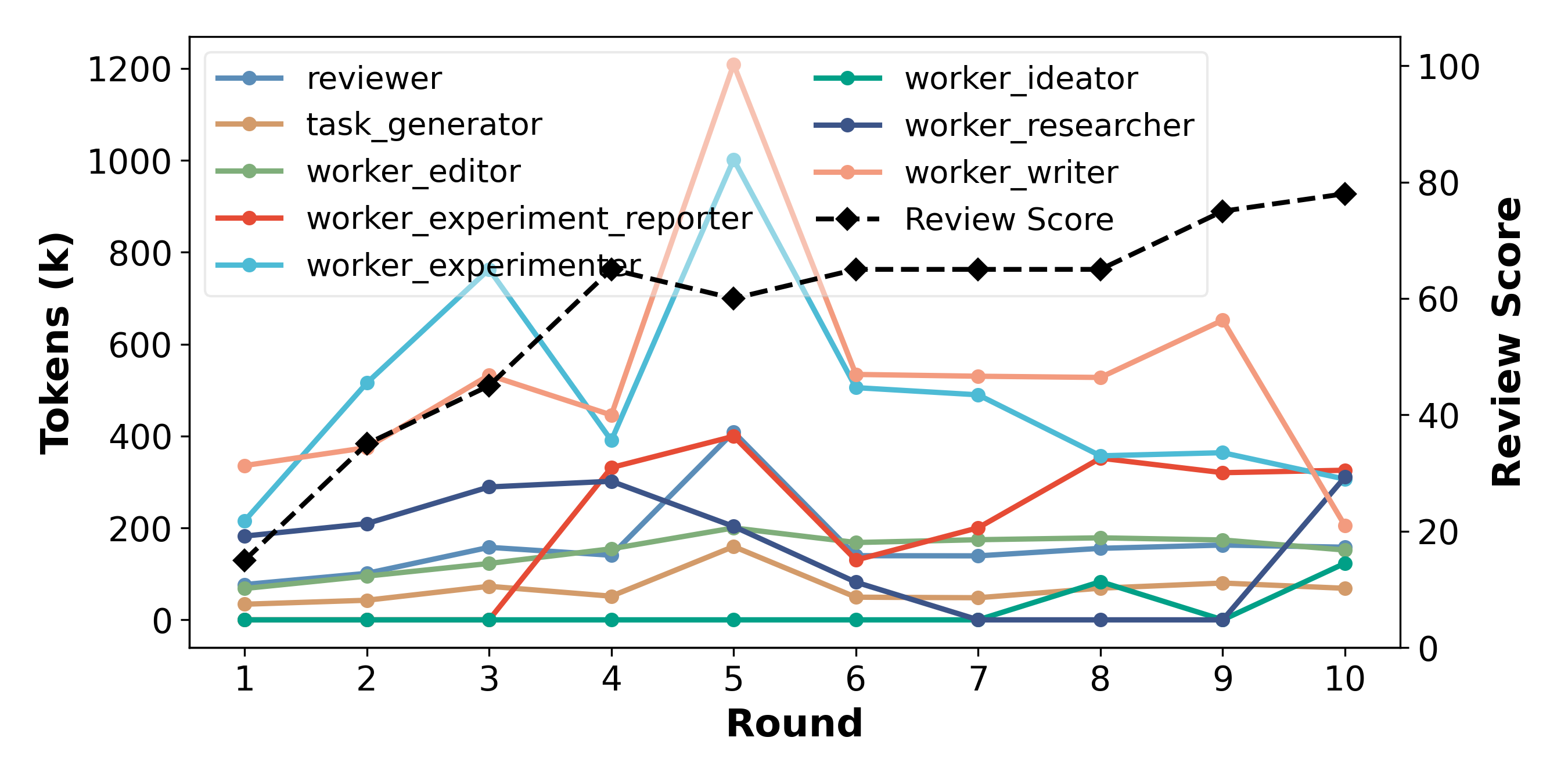}
  \caption{Illustrative per-round trace of task volume and token usage in the replay workflow.}
  \label{fig:lss-exp2-iter}
\end{figure}

We record per-round task counts and token traces for each type of agent at every round in Figure~\ref{fig:lss-exp2-iter} (DeepSeek API). For typical research, the experimental component often consumes far more tokens than other parts, and exploration can potentially continue indefinitely. To manage this, we reduced the activity of the Experiment Agent, allowing only a single round of basic experimentation. Even so, the Experiment Agent still consumed the most tokens. Throughout the run, LSS dynamically generated 23 skills. In this specific evaluation, we simplified the process by using \texttt{task.md} as a View to trigger Worker execution and categorized token usage by task type. This bypassed the Semantic Router, although a router would normally be required to dispatch tasks to existing Workers. Figure~\ref{fig:lss-exp2-iter} also presents the AI reviewer's subjective evaluation scores for each round, including both the generated manuscript sections and the image prompts.

\section{Conclusion}\label{sec:conclusion}

Multi-agent systems increasingly behave like Loosely-Structured Software. It is sometimes assumed that as AI models become capable of end-to-end autonomous execution, the need for system engineering will shrink. In practice, a transitional period may emerge: higher autonomy makes free-form interactions more open-ended, which can increase system-level entropy. In the LSS paradigm, the value of engineering is therefore less about prescribing step-by-step task logic and more about designing the macro-level ``physics''---governance mechanisms for context isolation, structural bindings, and safe self-evolution---so that a system of capable agents remains stable under repeated use. Moreover, these engineering design experiences also benefit AI development itself. It also suggests a shift in the software ecosystem: if most executable behavior is synthesized \emph{at runtime}, the volume of ``generated code'' may eventually dwarf the amount of code written ahead of time. This challenges today's software development and distribution models and raises unresolved questions about intellectual property and ownership.

\section{AI Use Statement}\label{sec:ai-use}
The authors acknowledge that generative AI contributed to this manuscript. GPT-5.2 (OpenAI) supported refinement of languages, while Nano Banana Pro 2 (Google) assisted with some image generation. The automated research environment proposed in this paper in section~\ref{sec:workflows} helped to explore new ideas and perform preliminary validation. All AI-generated content was critically reviewed, edited, and approved by the authors, who retain full responsibility for the manuscript's integrity and accuracy.

\bibliographystyle{ACM-Reference-Format}
\bibliography{sample-base}

@article{mei2025contextengineering,
  title={A survey of context engineering for large language models},
  author={Mei, Lingrui and Yao, Jiayu and Ge, Yuyao and Wang, Yiwei and Bi, Baolong and Cai, Yujun and Liu, Jiazhi and Li, Mingyu and Li, Zhong-Zhi and Zhang, Duzhen and others},
  journal={arXiv preprint arXiv:2507.13334},
  year={2025}
}

@article{liu2023lostmiddlelanguagemodels,
  title={Lost in the middle: How language models use long contexts},
  author={Liu, Nelson F and Lin, Kevin and Hewitt, John and Paranjape, Ashwin and Bevilacqua, Michele and Petroni, Fabio and Liang, Percy},
  journal={Transactions of the association for computational linguistics},
  volume={12},
  pages={157--173},
  year={2024}
}

@article{lewis2020rag,
  title={Retrieval-augmented generation for knowledge-intensive nlp tasks},
  author={Lewis, Patrick and Perez, Ethan and Piktus, Aleksandra and Petroni, Fabio and Karpukhin, Vladimir and Goyal, Naman and K{\"u}ttler, Heinrich and Lewis, Mike and Yih, Wen-tau and Rockt{\"a}schel, Tim and others},
  journal={Advances in neural information processing systems},
  volume={33},
  pages={9459--9474},
  year={2020}
}

@inproceedings{asai2023self,
  title={Self-rag: Learning to retrieve, generate, and critique through self-reflection},
  author={Asai, Akari and Wu, Zeqiu and Wang, Yizhong and Sil, Avirup and Hajishirzi, Hannaneh},
  booktitle={The Twelfth International Conference on Learning Representations},
  year={2023}
}

@article{karpas2022mrkl,
  title={MRKL Systems: A modular, neuro-symbolic architecture that combines large language models, external knowledge sources and discrete reasoning},
  author={Karpas, Ehud and Abend, Omri and Belinkov, Yonatan and Lenz, Barak and Lieber, Opher and Ratner, Nir and Shoham, Yoav and Bata, Hofit and Levine, Yoav and Leyton-Brown, Kevin and others},
  journal={arXiv preprint arXiv:2205.00445},
  year={2022}
}

@inproceedings{yao2023react,
  title={React: Synergizing reasoning and acting in language models},
  author={Yao, Shunyu and Zhao, Jeffrey and Yu, Dian and Du, Nan and Shafran, Izhak and Narasimhan, Karthik R and Cao, Yuan},
  booktitle={The eleventh international conference on learning representations},
  year={2022}
}

@article{schick2023toolformer,
  title={Toolformer: Language models can teach themselves to use tools},
  author={Schick, Timo and Dwivedi-Yu, Jane and Dess{\`\i}, Roberto and Raileanu, Roberta and Lomeli, Maria and Hambro, Eric and Zettlemoyer, Luke and Cancedda, Nicola and Scialom, Thomas},
  journal={Advances in neural information processing systems},
  volume={36},
  pages={68539--68551},
  year={2023}
}

@article{patil2023gorilla,
  title={Gorilla: Large language model connected with massive apis},
  author={Patil, Shishir G and Zhang, Tianjun and Wang, Xin and Gonzalez, Joseph E},
  journal={Advances in Neural Information Processing Systems},
  volume={37},
  pages={126544--126565},
  year={2024}
}

@article{qin2023toolllm,
  title={Toolllm: Facilitating large language models to master 16000+ real-world apis},
  author={Qin, Yujia and Liang, Shihao and Ye, Yining and Zhu, Kunlun and Yan, Lan and Lu, Yaxi and Lin, Yankai and Cong, Xin and Tang, Xiangru and Qian, Bill and others},
  journal={arXiv preprint arXiv:2307.16789},
  year={2023}
}

@inproceedings{wu2023autogen,
  title={Autogen: Enabling next-gen LLM applications via multi-agent conversations},
  author={Wu, Qingyun and Bansal, Gagan and Zhang, Jieyu and Wu, Yiran and Li, Beibin and Zhu, Erkang and Jiang, Li and Zhang, Xiaoyun and Zhang, Shaokun and Liu, Jiale and others},
  booktitle={First conference on language modeling},
  year={2024}
}

@article{packer2023memgpt,
  title={MemGPT: towards LLMs as operating systems.},
  author={Packer, Charles and Fang, Vivian and Patil, Shishir\_G and Lin, Kevin and Wooders, Sarah and Gonzalez, Joseph\_E},
  year={2023},
  publisher={ArXiv}
}

@inproceedings{greshake2023indirectprompt,
  title={Not what you've signed up for: Compromising real-world llm-integrated applications with indirect prompt injection},
  author={Greshake, Kai and Abdelnabi, Sahar and Mishra, Shailesh and Endres, Christoph and Holz, Thorsten and Fritz, Mario},
  booktitle={Proceedings of the 16th ACM workshop on artificial intelligence and security},
  pages={79--90},
  year={2023}
}

@article{liu2023agentbench,
  title={Agentbench: Evaluating llms as agents},
  author={Liu, Xiao and Yu, Hao and Zhang, Hanchen and Xu, Yifan and Lei, Xuanyu and Lai, Hanyu and Gu, Yu and Ding, Hangliang and Men, Kaiwen and Yang, Kejuan and others},
  journal={arXiv preprint arXiv:2308.03688},
  year={2023}
}

@inproceedings{qian2024chatdev,
  title={Chatdev: Communicative agents for software development},
  author={Qian, Chen and Liu, Wei and Liu, Hongzhang and Chen, Nuo and Dang, Yufan and Li, Jiahao and Yang, Cheng and Chen, Weize and Su, Yusheng and Cong, Xin and others},
  booktitle={Proceedings of the 62nd annual meeting of the association for computational linguistics (volume 1: Long papers)},
  pages={15174--15186},
  year={2024}
}

@inproceedings{hong2023metagpt,
  title={MetaGPT: Meta programming for a multi-agent collaborative framework},
  author={Hong, Sirui and Zhuge, Mingchen and Chen, Jonathan and Zheng, Xiawu and Cheng, Yuheng and Wang, Jinlin and Zhang, Ceyao and Wang, Zili and Yau, Steven Ka Shing and Lin, Zijuan and others},
  booktitle={The twelfth international conference on learning representations},
  year={2023}
}

@article{xia2024eddops,
  title={Evaluation-Driven Development and Operations of LLM Agents: A Process Model and Reference Architecture},
  author={Xia, Boming and Lu, Qinghua and Zhu, Liming and Xing, Zhenchang and Zhao, Dehai and Zhang, Hao},
  journal={arXiv preprint arXiv:2411.13768},
  year={2024}
}

@article{wen2025instructdetector,
  title={Defending against indirect prompt injection by instruction detection},
  author={Wen, Tongyu and Wang, Chenglong and Yang, Xiyuan and Tang, Haoyu and Xie, Yueqi and Lyu, Lingjuan and Dou, Zhicheng and Wu, Fangzhao},
  journal={arXiv preprint arXiv:2505.06311},
  volume={2},
  year={2025}
}

@article{wang2024survey,
  title={A survey on large language model based autonomous agents},
  author={Wang, Lei and Ma, Chen and Feng, Xueyang and Zhang, Zeyu and Yang, Hao and Zhang, Jingsen and Chen, Zhiyuan and Tang, Jiakai and Chen, Xu and Lin, Yankai and others},
  journal={Frontiers of Computer Science},
  volume={18},
  number={6},
  pages={186345},
  year={2024},
  publisher={Springer}
}

@article{hou2025model,
  title={Model context protocol (mcp): Landscape, security threats, and future research directions},
  author={Hou, Xinyi and Zhao, Yanjie and Wang, Shenao and Wang, Haoyu},
  journal={ACM Transactions on Software Engineering and Methodology},
  year={2025},
  publisher={ACM New York, NY}
}

@misc{mcp2025spec,
  title = {Specification - Model Context Protocol},
  author = {{Model Context Protocol Contributors}},
  year = {2025},
  url = {https://modelcontextprotocol.io/specification/2025-11-25},
  note = {Accessed: 2026-02-26}
}

@misc{a2a2025spec,
  title = {A2A Protocol Documentation},
  author = {{A2A Project Contributors}},
  year = {2025},
  url = {https://a2a-protocol.org/latest/},
}

@article{parnas1972criteria,
  title={On the criteria to be used in decomposing systems into modules},
  author={Parnas, David Lorge},
  journal={Communications of the ACM},
  volume={15},
  number={12},
  pages={1053--1058},
  year={1972},
  publisher={ACm New York, NY, USA}
}

@article{dragoni2017microservices,
  title={Microservices: yesterday, today, and tomorrow},
  author={Dragoni, Nicola and Giallorenzo, Saverio and Lafuente, Alberto Lluch and Mazzara, Manuel and Montesi, Fabrizio and Mustafin, Ruslan and Safina, Larisa},
  journal={Present and ulterior software engineering},
  pages={195--216},
  year={2017},
  publisher={Springer}
}

@article{stevens1974structureddesign,
  title={Structured design},
  author={Stevens, Wayne P. and Myers, Glenford J. and Constantine, Larry L.},
  journal={IBM systems journal},
  volume={13},
  number={2},
  pages={115--139},
  year={1974},
  publisher={IBM}
}

@book{kiczales1991amop,
  title={The art of the metaobject protocol},
  author={Kiczales, Gregor and Des Rivieres, Jim and Bobrow, Daniel G},
  year={1991},
  publisher={MIT press}
}

@inproceedings{marr2015zerooverhead,
  title={Zero-overhead metaprogramming: Reflection and metaobject protocols fast and without compromises},
  author={Marr, Stefan and Seaton, Chris and Ducasse, St{\'e}phane},
  booktitle={Proceedings of the 36th ACM SIGPLAN Conference on Programming Language Design and Implementation},
  pages={545--554},
  year={2015}
}

@article{zhong2023microservicecoupling,
  title={On measuring coupling between microservices},
  author={Zhong, Chenxing and Zhang, He and Li, Chao and Huang, Huang and Feitosa, Daniel},
  journal={Journal of Systems and Software},
  volume={200},
  pages={111670},
  year={2023},
  publisher={Elsevier}
}

@article{lercher2024microserviceapievolution,
  title={Microservice API evolution in practice: A study on strategies and challenges},
  author={Lercher, Alexander and Glock, Johann and Macho, Christian and Pinzger, Martin},
  journal={Journal of Systems and Software},
  volume={215},
  pages={112110},
  year={2024},
  publisher={Elsevier}
}

@misc{chase2022langchain,
  title = {LangChain},
  author = {Harrison Chase and LangChain Contributors},
  year = {2022},
  url = {https://github.com/langchain-ai/langchain},
}

@article{yang2023autogpt,
  title={Auto-gpt for online decision making: Benchmarks and additional opinions},
  author={Yang, Hui and Yue, Sifu and He, Yunzhong},
  journal={arXiv preprint arXiv:2306.02224},
  year={2023}
}

@book{martin2002agile,
  title={Agile software development, principles, patterns, and practices},
  author={Martin, Robert and Martin, Robert C},
  year={2013},
  publisher={BoD--Books on Demand}
}

@book{gamma1994designpatterns,
  title={Design patterns: elements of reusable object-oriented software},
  author={Gamma, Erich},
  year={1995},
  publisher={Pearson Education India}
}

@misc{openclaw2026,
  title = {OpenClaw: A Local Gateway for High-Reliability AI Agents},
  author = {{OpenClaw Team}},
  year = {2026},
  url = {https://openclaw.ai}
}

@misc{nanoclaw2026,
  title = {NanoClaw: Secure Personal AI Agent},
  author = {{NanoClaw Project}},
  year = {2026},
  url = {https://nanoclaw.dev/},
}

@misc{openai2025practicalagents,
  title = {A Practical Guide to Building Agents},
  author = {{OpenAI}},
  year = {2025},
  url = {https://cdn.openai.com/business-guides-and-resources/a-practical-guide-to-building-agents.pdf},
}

@misc{openai2026harnessengineering,
  title = {Harness Engineering: Leveraging Codex in an Agent-First World},
  author = {{OpenAI}},
  year = {2026},
  url = {https://openai.com/index/harness-engineering/},
}

@misc{fowler2026harnessengineering,
  title = {Harness Engineering},
  author = {Fowler, Martin},
  year = {2026},
  url = {https://martinfowler.com/articles/exploring-gen-ai/harness-engineering.html},
}

@article{khattab2024dspy,
  title={Dspy: Compiling declarative language model calls into self-improving pipelines},
  author={Khattab, Omar and Singhvi, Arnav and Maheshwari, Paridhi and Zhang, Zhiyuan and Santhanam, Keshav and Vardhamanan, Sri and Haq, Saiful and Sharma, Ashutosh and Joshi, Thomas T and Moazam, Hanna and others},
  journal={arXiv preprint arXiv:2310.03714},
  year={2023}
}

@inproceedings{park2023generativeagents,
  title={Generative agents: Interactive simulacra of human behavior},
  author={Park, Joon Sung and O'Brien, Joseph and Cai, Carrie Jun and Morris, Meredith Ringel and Liang, Percy and Bernstein, Michael S},
  booktitle={Proceedings of the 36th annual acm symposium on user interface software and technology},
  pages={1--22},
  year={2023}
}

@article{wang2023voyager,
  title={Voyager: An open-ended embodied agent with large language models},
  author={Wang, Guanzhi and Xie, Yuqi and Jiang, Yunfan and Mandlekar, Ajay and Xiao, Chaowei and Zhu, Yuke and Fan, Linxi and Anandkumar, Anima},
  journal={arXiv preprint arXiv:2305.16291},
  year={2023}
}

@article{shinn2023reflexion,
  title={Reflexion: Language agents with verbal reinforcement learning},
  author={Shinn, Noah and Cassano, Federico and Gopinath, Ashwin and Narasimhan, Karthik and Yao, Shunyu},
  journal={Advances in neural information processing systems},
  volume={36},
  pages={8634--8652},
  year={2023}
}

@article{zhong2023memorybank,
title={MemoryBank: Enhancing Large Language Models with Long-Term Memory},
volume={38}, 
number={17},
journal={Proceedings of the AAAI Conference on Artificial Intelligence},
author={Zhong, Wanjun and Guo, Lianghong and Gao, Qiqi and Ye, He and Wang, Yanlin},
year={2024},
month={Mar.},
pages={19724-19731}
}

@article{xu2025amem,
  title={A-mem: Agentic memory for llm agents},
  author={Xu, Wujiang and Liang, Zujie and Mei, Kai and Gao, Hang and Tan, Juntao and Zhang, Yongfeng},
  journal={arXiv preprint arXiv:2502.12110},
  year={2025}
}

@article{han2025atp,
  title={Alignment Tipping Process: How Self-Evolution Pushes LLM Agents Off the Rails},
  author={Han, Siwei and Xiong, Kaiwen and Liu, Jiaqi and Ye, Xinyu and Su, Yaofeng and Duan, Wenbo and Liu, Xinyuan and Xie, Cihang and Bansal, Mohit and Ding, Mingyu and others},
  journal={arXiv preprint arXiv:2510.04860},
  year={2025}
}

@article{rath2026agentdrift,
  title={Agent Drift: Quantifying Behavioral Degradation in Multi-Agent LLM Systems Over Extended Interactions},
  author={Rath, Abhishek},
  journal={arXiv preprint arXiv:2601.04170},
  year={2026}
}

@article{madaan2023selfrefine,
  title={Self-refine: Iterative refinement with self-feedback},
  author={Madaan, Aman and Tandon, Niket and Gupta, Prakhar and Hallinan, Skyler and Gao, Luyu and Wiegreffe, Sarah and Alon, Uri and Dziri, Nouha and Prabhumoye, Shrimai and Yang, Yiming and others},
  journal={Advances in neural information processing systems},
  volume={36},
  pages={46534--46594},
  year={2023}
}

@book{hebb1949organization,
  title={The organization of behavior: A neuropsychological theory},
  author={Hebb, Donald Olding},
  year={2005},
  publisher={Psychology press}
}

@article{christiano2017preferences,
  title={Deep reinforcement learning from human preferences},
  author={Christiano, Paul F and Leike, Jan and Brown, Tom and Martic, Miljan and Legg, Shane and Amodei, Dario},
  journal={Advances in neural information processing systems},
  volume={30},
  year={2017}
}

@article{brown2020fewshot,
  title={Language models are few-shot learners},
  author={Brown, Tom and Mann, Benjamin and Ryder, Nick and Subbiah, Melanie and Kaplan, Jared D and Dhariwal, Prafulla and Neelakantan, Arvind and Shyam, Pranav and Sastry, Girish and Askell, Amanda and others},
  journal={Advances in neural information processing systems},
  volume={33},
  pages={1877--1901},
  year={2020}
}

@inproceedings{reynolds2021promptprogramming,
  title={Prompt programming for large language models: Beyond the few-shot paradigm},
  author={Reynolds, Laria and McDonell, Kyle},
  booktitle={Extended abstracts of the 2021 CHI conference on human factors in computing systems},
  pages={1--7},
  year={2021}
}

@article{wei2022chainofthought,
  title={Chain-of-thought prompting elicits reasoning in large language models},
  author={Wei, Jason and Wang, Xuezhi and Schuurmans, Dale and Bosma, Maarten and Xia, Fei and Chi, Ed and Le, Quoc V and Zhou, Denny and others},
  journal={Advances in neural information processing systems},
  volume={35},
  pages={24824--24837},
  year={2022}
}

@article{wang2022selfconsistency,
  title={Self-Consistency Improves Chain of Thought Reasoning in Language Models},
  author={Xuezhi Wang and Jason Wei and Dale Schuurmans and Quoc Le and Ed Chi and Sharan Narang and Aakanksha Chowdhery and Denny Zhou},
  journal={arXiv preprint arXiv:2203.11171},
  year={2023}
}

@article{yao2023treeofthoughts,
  title={Tree of thoughts: Deliberate problem solving with large language models},
  author={Yao, Shunyu and Yu, Dian and Zhao, Jeffrey and Shafran, Izhak and Griffiths, Tom and Cao, Yuan and Narasimhan, Karthik},
  journal={Advances in neural information processing systems},
  volume={36},
  pages={11809--11822},
  year={2023}
}

@article{kim2025towards,
  title={Towards a science of scaling agent systems},
  author={Kim, Yubin and Gu, Ken and Park, Chanwoo and Park, Chunjong and Schmidgall, Samuel and Heydari, A Ali and Yan, Yao and Zhang, Zhihan and Zhuang, Yuchen and Malhotra, Mark and others},
  journal={arXiv preprint arXiv:2512.08296},
  year={2025}
}

@article{ojewale2026audittrailsaccountabilitylarge,
  title={Audit Trails for Accountability in Large Language Models},
  author={Ojewale, Victor and Suresh, Harini and Venkatasubramanian, Suresh},
  journal={arXiv preprint arXiv:2601.20727},
  year={2026}
}

@inproceedings{souza2025llmagentsinteractiveworkflow,
  title={LLM Agents for Interactive Workflow Provenance: Reference Architecture and Evaluation Methodology},
  author={Souza, Renan and Poteet, Timothy and Etz, Brian and Rosendo, Daniel and Gueroudji, Amal and Shin, Woong and Balaprakash, Prasanna and Ferreira da Silva, Rafael},
  booktitle={Proceedings of the SC'25 Workshops of the International Conference for High Performance Computing, Networking, Storage and Analysis},
  pages={2257--2268},
  year={2025}
}

@inproceedings{errica2025did,
  title={What did i do wrong? quantifying llms’ sensitivity and consistency to prompt engineering},
  author={Errica, Federico and Sanvito, Davide and Siracusano, Giuseppe and Bifulco, Roberto},
  booktitle={Proceedings of the 2025 Conference of the Nations of the Americas Chapter of the Association for Computational Linguistics: Human Language Technologies (Volume 1: Long Papers)},
  pages={1543--1558},
  year={2025}
}

@misc{liu2023repobench,
      title={RepoBench: Benchmarking Repository-Level Code Auto-Completion Systems}, 
      author={Tianyang Liu and Canwen Xu and Julian McAuley},
      year={2023},
      eprint={2306.03091},
      archivePrefix={arXiv},
      primaryClass={cs.CL}
}

\end{sloppypar}
\end{document}